\documentclass[journal,10pt]{IEEEtran}
\usepackage{epsfig,amsmath,amssymb,epsf,comment,setspace,cancel,graphicx,multirow,subfigure,eqnarray}
\usepackage{dsfont,mathrsfs}
\usepackage{xcolor}
\usepackage[mathscr]{euscript}
\usepackage{enumerate}
\DeclareGraphicsExtensions{.eps}
\usepackage{epstopdf}
\usepackage{cite}
\usepackage{pifont}
\usepackage[english]{babel}
\usepackage[utf8]{inputenc}
\usepackage{algorithm}
\usepackage{stfloats}
\usepackage[noend]{algpseudocode}
\usepackage{tikz}
\usepackage{bbm}
\usepackage{amsfonts}

\def\bff{{\mathbf{f}}}

\def\bq{{\mathbf{q}}}

\def\bw{{\mathbf{w}}}
\def\bx{{\mathbf{x}}}

\def\bQ{{\mathbf{Q}}}

\def\bX{{\mathbf{X}}}

\def\cH{{\mathcal{H}}}

\def\cP{{\mathcal{P}}}
\def\cQ{{\mathcal{Q}}}
\def\cR{{\mathcal{R}}}
\def\cS{{\mathcal{S}}}

\def\cX{{\mathcal{X}}}
\def\cY{{\mathcal{Y}}}

\def\argmin{\mathop{\mathrm{argmin}}}
\def\argmax{\mathop{\mathrm{argmax}}}

\def\b0{{\mathbf{0}}}

\def\sgn{\mbox{sgn}}

\definecolor{darkspringgreen}{rgb}{0.09, 0.45, 0.27}
\allowdisplaybreaks
\newcommand\numberthis{\addtocounter{equation}{1}\tag{\theequation}}
   
\newtheorem{theorem}{\textbf{{Theorem}}}
\newtheorem{lemma}{\textbf{{Lemma}}}
\newtheorem{definition}{Definition}
\newtheorem{prop}{\textbf{{Proposition}}}

\newtheorem{remark}{\textbf{{Remark}}}

\author{\IEEEauthorblockN{
Jing Guo,
Raghu G. Raj, \textit{Senior Member, IEEE},
David J. Love, \\ \textit{Fellow, IEEE}, and
Christopher G. Brinton, \textit{Senior Member, IEEE}
}\thanks{J. Guo, D. J. Love, and C. G. Brinton are with the School of Electrical and Computer Engineering, Purdue University, IN 47907 USA. email: richie89311@gmail.com; \{djlove, cgb\}\@purdue.edu}
\thanks{R. Raj is with the U.S. Naval Research Laboratory, DC 20375 USA. email: raghu.g.raj@gmail.com}}
\begin{document}
\title{\huge Nonparametric Decentralized Detection and Sparse Sensor Selection via Multi-Sensor Online Kernel Scalar Quantization}
\maketitle  
\IEEEdisplaynontitleabstractindextext
\IEEEpeerreviewmaketitle

\begin{abstract}
Signal classification problems arise in a wide variety of applications, and their  demand is only expected to grow.
In this paper, we focus on the wireless sensor network signal classification setting,  where each sensor forwards  quantized signals to a fusion center to be classified.
Our primary goal is to train a decision function and quantizers across the sensors to maximize the classification performance in an online manner.
Moreover, we are interested in sparse sensor selection using a marginalized weighted kernel approach to improve network resource efficiency by disabling less reliable sensors with minimal effect on classification performance.
To achieve our goals, we develop a multi-sensor online kernel scalar quantization (MSOKSQ) learning strategy that operates on the sensor outputs at the fusion center.
Our theoretical analysis reveals how the proposed algorithm affects the quantizers across the sensors.
Additionally,  we provide a convergence analysis of our online learning approach by studying its relationship to  batch learning.
We conduct numerical studies under different classification and sensor network settings which demonstrate the accuracy gains from optimizing different components of MSOKSQ  and robustness to reduction in the number of sensors selected.

\end{abstract} 
     
\section{Introduction}\label{sec:intro} 
Wireless sensor networks (WSNs), which leverage a set of spatially distributed  sensors to sample and  process  data from an environment, have been extensively studied in the past few decades \cite{Google,Tang,Guo,Ibrahim,Ordentlich, Wu, Avestimehr ,Yuksel, Lim, Sanderovich,Liang,Guoao, Nguyen, Wang12345,Koppel, Koppel1}.
A variety of WSN applications in, e.g., wireless communications
\cite{Google,Tang,Guo,Ibrahim,Ordentlich, Wu, Avestimehr,Yuksel, Lim, Sanderovich} and  radar  \cite{Liang} are increasingly having a significant impact our day-to-day lives. 
This widespread use of WSNs is driving research in a  number of  directions.  Of particular continuing interest is leveraging WSNs to provide  detection/estimation of one or more  underlying phenomena, specifically for signal classification tasks \cite{Nguyen, Wang12345, Koppel, Koppel1,Guoao}. 
%
{Several works have studied wireless signal classification and related problems in non-WSN settings as well, i.e., where the observations are centralized \cite{Sahay9542973,Han7438736}.}

In this paper, we focus on  decentralized detection   \cite{Viswanathan123,Nasipuri, Kassam, MIbrahim, Nguyen, Wang12345, Koppel, Koppel1, Guoao,Predd1,Predd2} exploiting a compress-and-forward sensor network \cite{Ordentlich, Wu, Avestimehr ,Yuksel, Lim, Sanderovich,Tang,Guo,Ibrahim, Guoao}.
In this setting, each sensor forwards information resulting from a  local decision rule to a fusion center for further processing and decision-making.
Decentralized detection problems can be divided into   parametric \cite{Nasipuri, Kassam, MIbrahim} and   non-parametric categories \cite{Nguyen, Wang12345,Koppel, Koppel1, Guoao,Predd1,Predd2}.
A parametric decentralized detection problem assumes that a fusion center knows or partially knows the joint distribution of $\bX$ (a random vector) and $Y$ (a random variable), denoted as $P(\bX,Y)$, where $\bX$ is the observation vector made by the sensors and $Y$ is the class label of $\bX$.
However, the need to maintain accurate estimates of $P(\bX,Y)$ could lead to  performance limitations and an increase in system complexity \cite{Nguyen,Wang12345}.
Thus,  non-parametric decentralized detection is a desirable alternative.


We are interested specifically in non-parametric decentralized detection for signal classification tasks.
In the past, authors have approached this topic by leveraging kernel-based learning techniques, e.g., \cite{Nguyen, Wang12345, Guoao, Koppel ,Koppel1}.  
Kernel methods are commonly employed to reduce computational complexity in problems that require mapping non-linearly separable datasets to higher dimensional spaces for decision learning \cite{Nguyen, Wang12345, Guoao, Koppel ,Koppel1}. 
The scenarios studied in \cite{Nguyen, Wang12345, Guoao} most closely align with our focus in this paper.  In these works, 
each sensor forwards a quantized observation based on a  quantization rule (corresponding to a local decision) to a fusion center.  
Then, the fusion center employs a decision function that maps from the space of quantized observations to the set of real numbers via the kernel being employed.
The output of the decision function is used to infer the corresponding signal classification class label.
To maximize classification performance, a joint optimization problem with respect to the quantizers (i.e., quantization rules) and the decision function is considered.
%

The methods in \cite{Nguyen, Wang12345} are based on batch learning (BL), where model training is conducted after full batches  of data  have been collected. For contemporary WSN applications which generate large volumes of observations in short time intervals, BL approaches potentially incur significant computation overhead. Additionally, the  detector/estimator learnt by the BL approach is highly dependent on the unknown input training sample distribution, i.e., $P(\bX,Y)$.
If the dataset distribution used for training is mismatched to the current WSN inputs (i.e., the testing dataset), which will happen when the environment evolves over time, the learning performance (e.g., classification accuracy) can be degraded significantly.

{In settings where information on the underlying environment exists, e.g., in terms of a finite state space, contemporary methods for modeling system evolution between states over time (e.g., Markov jump processes \cite{Cheng9552609,XIN2022126537}) could potentially be employed to capture these dynamics and enable the application of BL approaches within particular WSN states. As mentioned previously, we are interested in non-parametric approaches that do not require such assumptions. To address these challenges, one possibility is to consider an online learning (OL) approach to signal classification, where the fusion center adapts the classifier structure dynamically, i.e., each time a new data sample is observed by the sensors.}
%

In this paper, our primary goal is to develop an OL methodology for WSNs that jointly optimizes the quantizers and the decision function dynamically. Our theoretical analysis will also establish relationships between our OL approach and the BL method studied in  \cite{Nguyen}.    
In doing so, we will see how the proposed algorithm affects the quantization rules based on some computationally friendly marginalized kernels.
  
Our prior work \cite{Guoao} developed an OL methodology for signal classification based on measurements from a single sensor. One of the key differences here is the need to coordinate multiple sensors in a WSN. Specifically, it can be expected that quantization outputs will have varying contributions to the signal classifier at the fusion center, e.g.,  some sensors are naturally less reliable when they are deployed over a large area. 
To promote higher network operating efficiency, we are thus also interested in sensor selection  \cite{Wang12345,Silva,Goodman1}, where a certain number of lowest contributing sensors are disabled.
We  accomplish this within our OL framework by introducing a marginalized weighted kernel \cite{Wang12345} that allows us to assess sensor quality.
Compared to more general kernel-based OL methodologies, e.g., \cite{Kivinen, Koppel, Koppel1, Gentile, Dekel, Orabona,  Raj}, we consider the joint optimization of quantization, classification and sensor selection strategies.


\textbf{Summary of contributions.} The main contributions of this paper are summarized as follows:
\begin{itemize}
\item[•]  We introduce a WSN system model for    signal classification with sparse sensor selection in the OL setting.
{We formalize the instantaneous regularized risk for the OL setting in terms of the relevant WSN system parameters:} decision function weights, quantization rules,  and weighted parameter vectors for sensor selection.

\item[•]  
We develop the multi-sensor online kernel scalar quantization (MSOKSQ) algorithm to update our WSN parameters as new measurements are collected over time. Our generalized algorithm is based on alternating (sub-)-gradient methods that leverage the convexity of our risk function with respect to each of the three groups of WSN parameters.
{We present a specific implementation of our methodology based on a computationally friendly kernel.}


\item[•] We theoretically analyze the convergence behavior of our MSOKSQ algorithm, comparing the risk function to prior BL approaches \cite{Nguyen,Wang12345}.
We analyze the quantization rule updates across the sensors and obtain conditions on our system parameters under which the quantizers become (approximately) deterministic.  
    

\end{itemize}  

Through numerical evaluation on synthetic and real-world datasets, we demonstrate the impact of different WSN system components on MSOKSQ learning performance. {We find that these numerical results validate our theoretical results.}

The rest of the paper is organized as follows. Section II introduces the system model and basic background on kernels and decision functions. In Section III, we introduce  the  risk functions for the  BL and OL approaches. The MSOKSQ algorithm is developed based on the marginalized weighted kernel in Section IV. In Section V, we theoretically analyze the quantization rule updates and characterize the convergence behavior of our methodology. Numerical evaluation (including a  randomly generated dataset and the multi-class UCI Iris dataset \cite{UCI})  follows in Section VI. Finally, we conclude the paper and discuss future work.

   
\section{System Model and Preliminaries }
In this section, we first describe  the WSN system model (Section \ref{wsnsm}).
Then, we provide basic background on kernels and  decision functions  for our estimation model (Section \ref{kadf}).

\subsection{Wireless Sensor Network System Model}\label{wsnsm}
\begin{figure}[!t]
\centering
\includegraphics[width=0.45\textwidth]{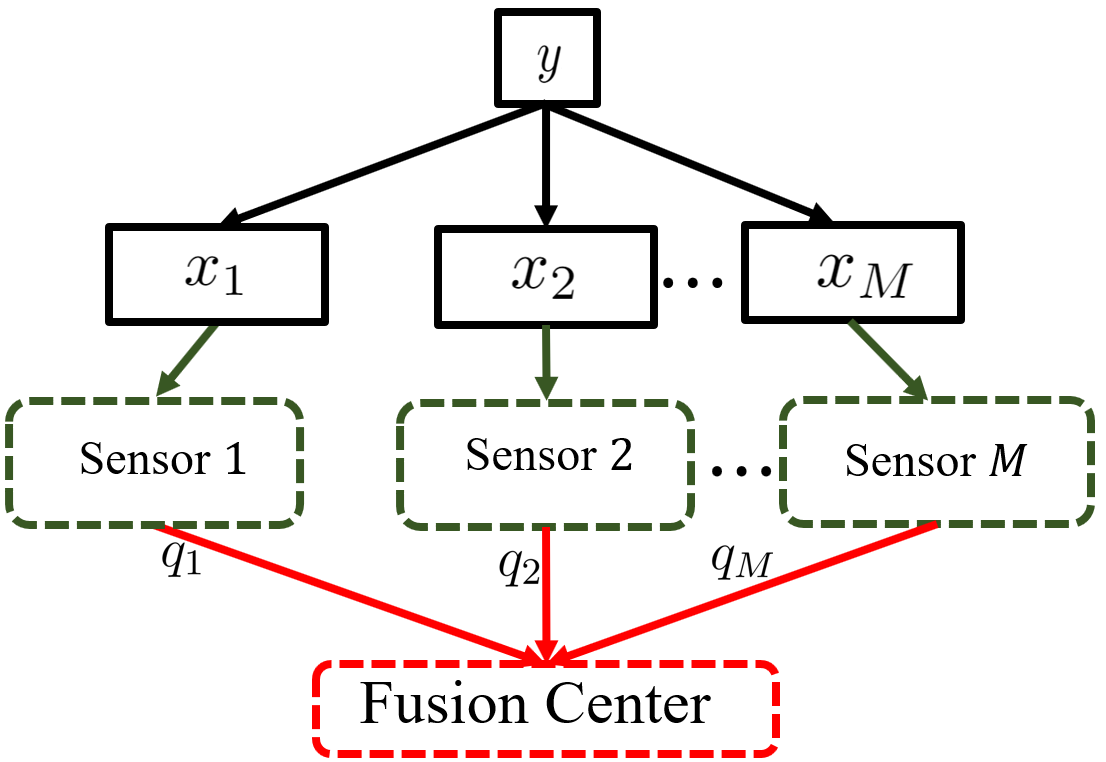}  
\caption{Illusion of the WSN system model we consider in this paper.}
\label{fig01}
\end{figure}

We consider  $M$ sensors connected to a common fusion center as shown in  Fig. \ref{fig01}.
Sensor $m$ is capable of measuring an observation $x_m \in \mathcal{X}_m$ of the environment, where $\mathcal{X}_m$ is finite collection of potential observations by sensor $m$.
 The fusion center must utilize the sensors to train a decision function to determine the class of a signal being observed by the sensors. We denote this class by  $y\in\mathcal{Y},$ 
representing  a  model for the target (dependent) variable of interest.
Given our focus on signal classification, 
$\cY$ is considered as a finite discrete set.
Although our proposed system model and methods are applicable to any multi-class problem, our presentation will often focus on the the binary classification case  for simplicity, i.e., $\cY\in \{-1,1\} $.
      
We assume perfect synchronization among sensors.
The primary functions of the fusion center are: i) training a decision function  chosen from a function space $\cH$ to determine the signal class  and ii) determining a quantization
rule to be employed by each sensor’s quantizer in order to
maximize classification performance.
To accommodate sensor selection, we consider a third function: iii) utilizing $M'$ of $M$ quantized observations, where $1\leq M'\leq M$, which can promote network resource efficiency through disabling lower quality sensors.

Ideally, the fusion center and quantizers would have access to  $P(\bX,Y)$ of the sensor observations and the signal class. Here, $Y\in \cY$ is a random  variable for the signal class $y$ and $\bX = [X_1,X_2,\ldots,X_M]^T$ is the random vector for observations across the sensors $\bx$. 
However,
as discussed in Section \ref{sec:intro}, this distribution is often unknown and difficult to estimate.
To overcome this lack of knowledge, we consider non-parametric decentralized detection based on a sequence of training samples $\cS = (s_n)_{n=1}^N$ such that $s_n =(\bx_n,y_n),$
 $\mathbf{x}_n = [x_{1,n}, \ldots, x_{M,n}] \in \mathcal{X}$ is the vector of observations over sensors at time $n$, $\mathcal{X} = \mathcal{X}_1 \times \cdots \times \mathcal{X}_M$, and $y_n \in \mathcal{Y}$. 
 {In the OL setting, this positive integer $N$ denotes the number of time instances $n = 1,...,N$ under consideration, where one sample arrives at each time instance.
 Note that it is usually treated as the total number of samples in the BL setting, e.g.,  \cite{Nguyen, Wang12345}.}

We let $q_{m,n} \in \mathcal{Q}_m$ denote sensor $m$'s quantized version of $x_{m,n}$ at time $n$. Across the sensors, we consider the vector $\bq_{n} = [q_{1,n}  \cdots q_{M,n}]^T \in \cQ$, where $\cQ = \mathcal{Q}_1\times \cdots\times \mathcal{Q}_M$. 
We assume $\cQ_m$ and $\cQ$ are   finite sets such that $|\cQ_m| \leq |\cQ|$, $|\cQ_m| \leq |\cX_m|$, and $|\cQ | \leq |\cX|$, for all $m$.  
Note that $|\cX|$ denotes the cardinality of the set $\cX$. $q_{m,n}$ is produced 
based on the quantization rule $P_{m,n}(q_{m,n}  | x_{m,n} )$ of sensor $m$ at time $n$.      
We expect in practice that $P_{m,n}(q_{m,n}|x_{m,n})$ for a particular value of $x_{m,n} \in \cX_m$ will be unimodal: Given $x_{m,n}$, there would be one value of $q_{m,n}$ where the density is the highest, and it would taper off from there as $q_{m,n}$ is varied, e.g., as a Gaussian distribution.
According to the axioms of probability \cite{papoulis02}, these rules must follow $P_{m,n}(q_{m,n} | x_{m,n}) \in [0,1]$ and  
$\sum_{q_{m,n} \in \cQ_m} P_{m,n}(q_{m,n} | x_{m,n}) = 1$,  for all $q_{m,n} \in \cQ_m$ and  $x_{m,n} \in \cX_m$.
We posit conditional independence of the quantization rules at time $n$ (which confers computational efficiency to our algorithms) \cite{Nguyen, Wang12345}:
\begin{equation}\label{qr3}
\begin{array}{ll} 
P_n(\bq_n | \bx_n)&= \displaystyle{\prod_{m=1}^M} P_{m,n}(q_{m,n}|x_{m,n})
\end{array}  
\end{equation}
for all possible $\bq_n$ and $\bx_n$.
As a result, we can form a set $\cP$ of the feasible quantization rules across the sensors as
\begin{equation}\label{mathcalp}
\begin{array}{ll} 
\cP&= \Big\{P(\bQ|\bX):P(\bq | \bx) = \displaystyle{\prod_{m=1}^M P_{m}(q_{m}| x_m)}, \forall\bq\in\mathcal{Q},\\
&\forall\bx\in\mathcal{X}, \displaystyle{\sum_{q_m\in\mathcal{Q}_m}} P_m(q_m |x_m) = 1, P_m(q
_m|x_m) \geq 0  \Big\},
\end{array}   
\end{equation}
where $\bQ \in \cQ$ is a random  vector.

After quantization, the WSN forwards $\bq_n$ to the fusion center.
Similar to \cite{Nguyen, Wang12345, Guoao}, we assume  that the fusion center perfectly observes $\bq_n$, i.e.,  the channel does not introduce significant noise to the quantized measurements.
The fusion center aims to jointly optimize the decision function, quantization rules, and sensor selection, thereby controlling the behavior of the quantizers (i.e., {$P_n(\bq_n|\bx_n)$}) at each time $n$.

\subsection{Kernel and Decision Function}\label{kadf}

\subsubsection{Kernel}
We are interested in developing an OL algorithm  that  can work with both linearly and non-linearly separable models of the quantized data.
One  common approach to dealing with  non-linearly separable models is  to map the dataset to a high-dimensional 
space $\cH$ using a  mapping function ${\phi}(\cdot):\cQ \rightarrow \cH$\cite{Wang12345, Nguyen, Guoao, Kivinen, Koppel, Koppel1, Scholkopf}.
Generally, this mapping function is not unique.   
An algorithm can then be developed based on  the inner product  between points, i.e., $\langle{\phi}(\bq^{d}),
{\phi}(\bq^{d'})
\rangle \in \mathbb{R}$  for all possible quantization points $\bq^{d},\bq^{d'}\in\cQ$, for $d, d' = 1,2\ldots |\cQ|$, 
so that the problem is solved using one or more (approximately) linear decision boundaries \cite{Wang12345, Nguyen, Guoao, Kivinen, Koppel, Koppel1, Scholkopf}.
Although it can be computationally  challenging to construct ${\phi}(\cdot)$, with kernel methods, we work with $\langle{\phi}(\bq^{d}),
{\phi}(\bq^{d'})\rangle$ instead, which is much easier to construct \cite{Scholkopf}.  
\begin{definition}\label{df1}
\textit{A real-valued symmetric function $k(\cdot,\cdot):\mathcal{Q}\times \mathcal{Q}\rightarrow\mathbb{R}$ is a kernel if 
the matrix $\mathbf{K} = [K_{d,d'}] \in \mathbb{R}^{|\mathcal{Q}| \times |\mathcal{Q}|}, K_{d,d'} = k(\bq^{d}, \bq^{d'})$ for $d, d' = 1,2...,|\cQ|$ is positive semidefinite, i.e., a Gram matrix.} 
\end{definition}

We can define the mapping function based on a kernel according to ${\phi}(\bq^{d}) = k(\bq^{d}, \cdot)$ so that 
$\langle  k(\bq^{d}, \cdot),  k(\bq^{d'}, \cdot)\rangle =   k(\bq^{d},\bq^{d'}).$
%
To facilitate sensor selection, we focus on weighted kernels in this paper \cite{Wang12345}. 
Specifically, a weighted kernel $k_{\bw}(\bq^{d},\bq^{d'}) $ is used to associate each sensor $m$ with a level of reliability using a weight parameter vector
$\bw = [w_1,\ldots ,w_M]^T$ that satisfies 
\begin{equation}\label{weightedparameterconstraint} 
    \| \bw\|_1 = M \text{ and } \bw \succeq \mathbf{0},
\end{equation} 
where $\bw \succeq \mathbf{0}$ indicates $w_{m} \geq 0$  for all $m$. 
  A larger weight $w_m$ indicates that the corresponding sensor is more reliable \cite{Wang12345}.
Furthermore, we consider sensor $m'$ to be disabled 
when $w_{m'}=0 $.
For this reason, the weighted kernel will be critical to our  algorithm development.
Note that the construction of the weighted kernel is not unique.   
An example weighted kernel is the weighted count kernel  described by \cite{Nguyen,Wang12345} \footnote[1]{Another example inspired by \cite{Nguyen} is the weighed linear kernel, e.g., $k_{\bw}(\bq^{d}, \bq^{d'}) =\sum_{m = 1}^{M}w_{m}^2 q_{m}^{d}  q_{m}^{d'}$.}
\begin{equation}\label{wkernel2}
k_{\bw}(\bq^{d}, \bq^{d'}) =\sum_{m = 1}^{M} w^2_{m} \mathds{1}( q_{m}^{d} = q_{m}^{d'}),
\end{equation}
where $\mathds{1}(\cdot)$ denotes the indicator function. $q_m^d$ and $q_m^{d'}$ are the $m$th element of $\bq^d$ and  $\bq^{d'}$, respectively. 
 



%
%

\subsubsection{Decision function}
The purpose of the decision function is to estimate/predict the class of the input samples from the quantized sensor measurements.
Formally, the decision function $f(\cdot)$ is defined as 
\begin{equation}\label{deec1}
f\big(\bq^{d} \big)=  \sum_{d'= 1}^{|\cQ|} \alpha_{d'} k_{\bw}(\bq^{d},\bq^{d'}),
\end{equation}  
where   $\alpha_{d'} \in \mathbb{R}$ for each quantization point $d'$. 
We assume that $\cH$ is a  reproducing kernel Hilbert space (RKHS).
A RKHS $\cH$ is one for which there is a kernel $k_{\bw}:\mathcal{Q} \times \mathcal{Q} \rightarrow \mathbb{R}$  such that  the following two properties hold:
\begin{itemize}
    \item $k_{\bw}$ has the reproducing property   
    \begin{equation}\label{dcfn1} 
f (\bq^{d}) = \langle \mathbf{f}, k_{\bw}(\bq^{d},\cdot)\rangle,  \text{for all } \mathbf{f} \in \mathcal{H} \text{ and } d. \end{equation}
    \item $\mathcal{H}$ is the closure of the space of all  $k_{\bw}(\bq^{d},\cdot)$, for all $d$.  
\end{itemize}   
We refer to $\bff \in \mathcal{H}$ as the weight of  $f(\cdot)$.
This means that \eqref{dcfn1} is the decision function induced by the mapping space $\cH$.
In addition, $\bff= \sum_{d= 1}^{|\cQ|} \alpha_{d } k_{\bw}(\cdot, \bq^{d })$ based on  \eqref{deec1} and \eqref{dcfn1}.   
   

 
\section{Estimation Model}\label{OA}
In this section, we formalize the signal classification problem  for the BL (Section III-A) and OL (Section III-B) settings.

 
\subsection{Prior Work: Batch Learning Problem Formulation}\label{opsu} 
BL has been employed for WSN signal classification in prior works \cite{Nguyen,Wang12345,Scholkopf}. In BL, training and estimation are conducted after the sample sequence $\mathcal{S}$ has been collected. This formulation will be useful to us in our subsequent development of the OL case.
A solution set for the BL approach can be denoted $\gamma = \big\{ \bff,  \{P(\bq_n |\bx_n) \}_{n = 1}^N , \bw, \forall\bq_n \in \cQ, \forall \bx_n  \in \cX  \big\}$.
The weight of the decision function $\bff$, the quantization rules across the sensors $P(\bq_n |\bx_n )$, 
 and the weight parameter vector $\bw$  are determined offline based on $\cS$, and thus they will not change over time as compared to the OL case.
{We use the subscript $n$ in the BL setting to denote the sample index.}
      
An empirical risk function was introduced by \cite{Nguyen,Wang12345} to obtain the optimal $\gamma$, which is given by
\begin{equation}\label{emprisk}
R_{\text{emp}}(\cS) = \sum_{n=1}^{N}\sum_{\mathbf{q}_n\in \mathcal{Q}}\frac{1}{N}\ell\left(f\big(\bq_n \big),y_{n}\right) P(\mathbf{q}_n| \bx_n),
\end{equation}      
where $\ell(\cdot, \cdot)$ is a loss function, which can either be a differentiable or non-differentiable convex function.
For example, the logistic loss is a differentiable function, which is
\begin{equation} 
\ell( f\big(\bq\big), y\big) := \log\Big(1+\exp\big\{-y f\left(\bq\right)\big\}\Big).
\end{equation}
On the other hand, the  soft margin loss function takes the following non-differentiable form \cite{Kivinen}:
\begin{equation}\label{hingeloss}
\ell\big( f\big(\bq \big), y\big) :=\max \big\{ 0, ~\rho -y f\big( \bq \big) \big\},
\end{equation}  
where $\rho >0$ is the margin parameter, $\bq \in \cQ$, and $y\in \cY$. \eqref{hingeloss} becomes the hinge loss function when $\rho = 1$, which is most notably used for the binary classification problem, e.g., \cite{Nguyen, Wang12345, Guoao, Scholkopf}.

To prevent model overfitting and perform sensor selection, a regularized form of the empirical risk function was introduced in \cite{Wang12345}, which is given by
 \begin{equation}\label{R2}
\begin{array}{ll}
 \bar{R}_{\text{reg}}(\cS) &=R_{\text{emp}}(\cS)  +\dfrac{\lambda_1\|\bff\|^2}{2}  + {\lambda_2} \| \bw \|_1,
\end{array} \end{equation}
{where $\|\cdot\|$ and $\| \cdot\|_1$ denote the $L_2$ and $L_1$ norms, respectively. 
This same choice of norms will also be used for the OL setting in Section \ref{ola}.}
The regularization parameters $\lambda_1, \lambda_2 \geq 0$ are treated as fixed \cite{Wang12345}. 
 $\lambda_2 \geq 0 $ controls how many reliable sensors are selected for the classification \cite{Wang12345}, with the $L_1 $ norm helping to enforce a sparse  solution  while retaining convexity.  
   
   
It is difficult to learn the desired parameters in \eqref{R2} based on an expectation of the loss function over $\bq_n$ for all $n$ \cite{Nguyen, Wang12345}.
Traditional upper-bounding techniques impose high computational complexity.
{To resolve this, \cite{Nguyen} and \cite{Wang12345} apply Jensen's inequality to work with the following lower bound of  \eqref{R2}:} 
\begin{equation}\label{sop1}
R_{\text{reg}}(\cS) =\sum_{n=1}^{N}\frac{\ell(\langle \bff, \Phi(\bx_n) \rangle ,y_{n})    }{N} +\frac{\lambda_1 \| \bff \|^2}{2}+ {\lambda_2}\|\bw\|_1,
\end{equation}
where $\Phi(\cdot) \in \cH$ is a marginalized mapping function based on \cite{Nguyen, Wang12345} for jointly solving $P(\bq_n|\bx_n)$ and $\bw_n$: 
\begin{equation}\label{mgk}
\Phi(\bx_n) = \sum_{\bq_n\in \mathcal{Q}}P(\bq_n| \bx_n) k_{\bw}(\bq_n,\cdot).
\end{equation} 
Note that \eqref{R2} and \eqref{sop1} are equal when the quantizers are  \textit{deterministic} \cite{Nguyen,Wang12345}, i.e.,
 \begin{equation}\label{determinsquantizationrules}
P_m(q_{m,n}|x_{m,n}) = \begin{cases}
1, & \text{if } q_{m,n} = q_m,\\
0, & \text{otherwise },
\end{cases} \text{ for some }q_m \in \cQ_m, 
 \end{equation}
for all $m$ and sample $n$. 
The optimal $\gamma$ for this deterministic case described in \eqref{determinsquantizationrules} can be derived for both \eqref{R2} and \eqref{sop1}, which has been studied in \cite{Nguyen, Wang12345}.
{Additionally, these works have shown that the lower bound risk (11) converges to (10) for a large training dataset.}
With this in hand, we move on formulating a risk function for the OL setting, which will build upon \eqref{sop1} for the BL case.
  
\begin{algorithm}[t!]
\caption{MSOKSQ algorithm  for signal classification.}\label{a1}
\begin{algorithmic}
\Procedure{MSOKSQ}{$\mathcal{S}$}  
\State Set system parameters: $N,M' > 0$  
\State Initialize  hyper-parameters: $ \eta_1,\eta_{1}^P, \eta_{1}^{\bw},   \lambda_1,\lambda_{2,1}> 0$
\State Initialize hyper-parameters: $\bff_1 = \mathbf{0}$, $\bw_1 = \mathbf{1}$, and \phantom{show}$P_{n(1)}(\bq_{n(1)} |\bx_{n(1)})$ based on Section \ref{updatequantizationrules}
\For { $n = 1:1:N-1$} 
\State Calculate $P_{m,n+1}(q_{m,n}| x_{m, n})$ using Theorem \ref{lema1} (or \phantom{show12}Proposition \ref{prop002})  \eqref{diffpqx1}, \eqref{diffpqx12}, and \eqref{pickjj}, given $\mathbf{w}_n$ and $\mathbf{f}_n$ 
\State Update $P_{n+1}(\bq_n | \bx_{n})$ using \eqref{qr3}
\State Calculate $w_{m, n+1}$ using Proposition \ref{prop001}, \eqref{bw1}, and \phantom{show12}\eqref{lambda2n}, given $\mathbf{f}_n$ and $P_{n+1}(\mathbf{q}_n | \mathbf{x}_n)$
\State Update  $\mathbf{w}_{n+1}$ based on $w_{m,n}$
\State Update $\bff_{n+1}$ using \eqref{wofatn}, given $\mathbf{w}_{n+1}$ and \phantom{show12}$P_{n+1}(\mathbf{q}_n | \mathbf{x}_n)$
\EndFor  
\State \textbf{return} $(\gamma_{n})_{n=1}^{N}$
\EndProcedure{\textbf{end procedure}}
\end{algorithmic}
\end{algorithm}

\subsection{Online Learning Problem Formulation}\label{ola}
One of the main contributions of this paper is to solve the decentralized signal classification problem in the online setting.
Specifically, we need to update the weight of the decision function $\bff_n \in \cH$, the quantization rules across the  sensors $P_{n}(\bq_{n}|\bx_{n}) \in \cP$, and the weighted parameter vector $\bw_n$ satisfying \eqref{weightedparameterconstraint} at each time $n$, i.e., whenever a new sample is observed.
We will use  $(\gamma_n = \{\bff_n, P_n(\bq_n|\bx_{n}),\bw_n  \})_{n = 1}^{N}$ as a solution set sequence across the sensors over time $n$.   

For the OL setting, we formulate the following instantaneous regularized risk function, inspired by  \cite{Wang12345,Kivinen}:
\begin{equation}\label{rmrk}
R(s_n) = \ell\big(\langle \bff_n, \Phi_n(\bx_n) \rangle ,y_{n}\big) +\dfrac{\lambda_1 \|\bff_n\|^2}{2} + {\lambda_{2,n}}\|\bw_n\|_1,
\end{equation}    
which is an  approximation of $R_{\text{reg}}(\cS)$ in \eqref{sop1} based on the sample  at  time $n$.
As before, $\lambda_1> 0 $ and $\lambda_{2,n}\geq 0$ are regularization terms.
Note that $\lambda_{2,n}$ controls how many sensors are disabled which may evolve  over time.
The modified decision function at time $n$ is $\langle \bff_n, \Phi_n(\bx_n) \rangle $, inspired by \eqref{sop1}.
The marginalized mapping function $\Phi_{n}(\bx_n)$ at time $n$ is
\begin{equation}\label{mmftime}
    \Phi_{n}(\bx_n)  = \displaystyle{\sum_{\bq_n\in \mathcal{Q}}}P_n(\bq_n| \bx_n)k_{\bw_{\cdot, n}}(\cdot,\bq_n),
\end{equation}
Compared to \eqref{mgk}, $k_{\bw_{\cdot, n}}(\cdot,\bq_n)$  can be  considered as the mapping function associated with $\bw_n$ at time $n$. $k_{\bw_{i,n}}(\bq_i, \bq_n)$ is the weighted kernel with respect to time $i$ and $n$.   
We define
$k_{\Phi_{i,n}}(\bx_i,\bx_n) = \langle \Phi_{i}(\bx_i) ,\Phi_{n}(\bx_n)\rangle$ as the  marginalized kernel with time $i$ and $n$ for developing an algorithm to solve \eqref{rmrk}:
\begin{equation}\label{normphi1}
k_{\Phi_{i,n}}(\bx_i,\bx_n) = \sum_{\bq_i, \bq_n \in \cQ}P_i(\bq_i|\bx_i)P_n(\bq_n| \bx_n)k_{\bw_{i,n}}(\bq_i, \bq_n).\end{equation} 

  

{The kernel calculation will be the most computationally intensive component of our methodology in Section IV. In general, (16) has a complexity of $O(|\mathcal{Q}|^N)$, which becomes prohibitive for a large number of training samples $N$. To address this, inspired by  \cite{Nguyen,Guoao, Wang12345}, we will consider a computationally friendly 
weighted marginalized kernel, namely, the
weighted count marginalized kernel \cite{Tsuda}:\footnote[2]{Another example is the marginalized weighed linear kernel with time $i$ and $n$, i.e., $
k_{\Phi_{i,n}}(\bx_i,\bx_n) ={\sum_{m = 1}^{M}w_{m,i}w_{m,n}\mathbb{E}[\phi(Q_m) |x_{m,i}]} \mathbb{E}[\phi(Q_m)|x_{m,n}]$, where $Q_m$ is $m$th element of the random vector $\bQ$.}
\begin{equation}\label{wkernel3}
\begin{array}{ll}
&k_{\Phi_{i,n}}(\bx_i,\bx_n)\\
&=\displaystyle{\sum_{m = 1}^{M}\sum_{q_{m,n},q_{m,i} \in \cQ_m}w_{m,i}w_{m,n}} 
 P_{m,i}(q_{m,i}|x_{m,i})\\ 
 &\phantom{sh} \times P_{m,n}(q_{m,n}|x_{m,n})
 \mathds{1}(q_{m,i} = q_{m,n},x_{m,i} = x_{m,n}). 
\end{array}\end{equation}
This reduces the kernel calculation complexity to $O(M^{N'})$, where $M \ll |\mathcal{Q}|$, and $N' \ll N$ is the number of time instances $i$ for which $q_{m,i} = q_{m,n}$ and $x_{m,i} = x_{m,n}$.}
{More generally, there are several ways to construct $k_{\Phi_{i,n}}(\bx_i,\bx_n) $, as long as the construction of $k_{\bw_{i,n}}(\bq_i, \bq_n)$ in \eqref{normphi1} follows Definition \ref{df1}.}

Compared to \cite{Wang12345}, an additional constraint $x_{m,i} = x_{m,n}$ has been added in the indicator $\mathds{1}(\cdot,\cdot)$,
for all $i = 1,2,\ldots,n$ because of the OL setting.  
This change ensures the each sensor quantizer becomes a(n) (approximate) deterministic quantizer in the limit.
{This is important because, similar to \eqref{sop1} for the BL case, \eqref{rmrk} is a lower bound on the risk function for the OL setting. Our theoretical analysis in Section V (Lemma \ref{lemma1}, Theorem \ref{proconnection}) will establish a relationship between the performance of these two lower bounds, and we will see that they converge once a large number of samples have been observed.}

\section{Online Learning Methodology}\label{OLA}
We will now develop our MSOKSQ methodology for the formulation in Section \ref{ola}. To do so, we note that while \eqref{rmrk} is non-convex with respect to $\gamma_n$, it is convex with respect to each of the variable sets $\mathbf{f}_n$, $P_n(\mathbf{q}_n | \mathbf{x}_n)$, and $\mathbf{w}_n$ individually while the other two are assumed to be fixed. Thus, we are motivated to develop a (sub-)gradient descent-based method that minimizes \eqref{rmrk} by iteratively updating each element in $(\gamma_n)_{n=1}^{N}$. 

In this paper, we  focus on the most general case of non-differential loss functions used in \eqref{rmrk}, where the \textit{sub-gradients} of elements in $\gamma_n$  prior to each update need to be derived.
We can extend the results of the algorithm to any differentiable loss function since the  sub-gradient of a differentiable loss function is its  gradient \cite{Scholkopf}.   

Procedurally, after we initialize $\bff_1$, $\bw_{1}$, and $P_1(\bq_1|\bx_1)$ for all possible $\bq_1 \in \cQ$ and $\bx_1 \in \cX$, we first must derive three updates. First is to derive $\bff_n$ given   $P_n(\bq_{n-1}|\bx_{n-1})$ and $\bw_{n-1}$ at time $n-1$ (Section \ref{bffn}), where $P_{n}(\bq_{n-1}|\bx_{n-1})$ is the updated version of $P_{n-1}(\bq_{n-1}|\bx_{n-1})$.
Second is to derive $P_{n+1}(\bq_n|\bx_{n})$ given   $\bff_{n}$ and $\bw_n$ at time $n$ (Section \ref{updatequantizationrules}).
Finally, we must obtain  $\bw_{n+1}$ given $\bff_n$ and $P_{n+1}(\bq_{n}|\bx_n)$ at time $n$ (Section \ref{weightedparameters}).  
 The resulting procedure is summarized in Algorithm \ref{a1}.

\subsection{Obtaining  weight of decision function $\bff_n$}\label{bffn}
Given $P_n(\bq_{n-1}|\bx_{n-1})$ and $\bw_n$, we update  $\bff_n$ 
 based on \eqref{rmrk} using
\begin{equation}\label{diffbff1}
\bff_{n} = \bff_{n-1} - \eta_{n-1} \partial_{\bff_{n-1}}R(s_{n-1}),
\end{equation}      
where $\eta_{n-1} \geq 0$ is the learning rate.   $\partial_{\bff_{n-1}}R(s_{n-1})$ is the sub-gradient of $R(s_{n-1})$ with respect to $\bff_{n-1}$, given by
\begin{equation}\label{dirrf} 
\begin{array}{ll}
&\partial_{\bff_{n-1}}R(s_{n-1}) \\
& \phantom{s}=\partial_{\langle \bff_{n-1}, \Phi_{n-1}(\bx_{n-1}) \rangle}\ell\big(\langle \bff_{n-1}, \tilde{\Phi}_{n-1}(\bx_{n-1}) \rangle,y_{n-1}\big)\\ 
&\phantom{ssss}\times \tilde{\Phi}_{n-1} (\bx_{n-1})+\lambda_1 \bff_{n-1}
\end{array}\end{equation}
using the chain rule, where $\tilde{\Phi}_{n-1}(\cdot)$ is the updated marginalized mapping function at time $n-1$
\begin{equation}\label{phipnpuls1} 
\tilde{\Phi}_{n-1} (\bx_{n-1}) = \sum_{\bq_{n-1}\in\cQ} P_{n}(\bq_{n-1} | \bx_{n-1})k_{\bw_{\cdot,{n-1}}}(\cdot,\bq_{n-1}).
\end{equation}
Initializing $\bff_1 = \mathbf{0}$,  we update $\bff_{n}$ as
\begin{equation}\label{wofatn}
\bff_{n} = \sum_{i = 1}^{n-1} \alpha_{i,n} \tilde{\Phi}_i(\bx_i),
\end{equation}
where  $\alpha_{i,n}$ is the coefficient at time $n$ of  $\tilde{\Phi}_i(\mathbf{x}_i)$, which is given by
\begin{equation}\label{alpha1234}
\alpha_{i,n} = \begin{cases}
-\eta_{i} \partial_{\langle \bff_{i}, \tilde{\Phi}_i(\bx_i) \rangle}\ell\big(\langle \bff_{i}, \tilde{\Phi}_i(\bx_i) \rangle, y_i\big), & \text{if }i = n-1,\\
(1-\eta_{n-1}\lambda_1)\alpha_{i,n-1},&\text{if } i < n-1. 
\end{cases}
\end{equation} 
 To form $\mathbf{f}_n$, the new coefficient  $\alpha_{n-1,n}$  is generated with $\tilde{\Phi}_{n-1} (\mathbf{x}_{n-1})$  based on the sub-gradient of $\ell(\cdot,\cdot)$, and the $\alpha_{i,n}$ on the previous $\tilde{\Phi}_{i}(\mathbf{x}_{i})$ are multiplied by $(1 - \eta_n \lambda_1)$ for $i = 1,2,\ldots,n-1$ as in \cite{Kivinen}. Followed by this, we can write the modified decision function, which is given by
\begin{equation}\label{weighttinstance}
\langle \bff_{n}, \Phi_n(\bx_n) \rangle= \sum_{i = 1}^{n-1}\alpha_{i,n}  k_{\Phi_{i,n}}(\bx_i, \bx_{n}).
\end{equation} 
   {We can see from \eqref{alpha1234} that we need $\lambda_1 < 1 / \eta_n$ for the updates to converge \cite{Kivinen}. Our theoretical analysis in Section V will not make additional assumptions on the value of $\lambda_1$, which can in general be chosen through cross-validation on a particular dataset.}
  



\subsection{Obtaining   $P_{n+1}(\bq_n|\bx_n)$ for all possible $\bq_n$ and  $\bx_n$} \label{updatequantizationrules}
 We derive $P_{n+1}(\bq_n |\bx_{n})\in \cP$ given $\bff_{n}$ and $\bw_n$ for all possible $\bq_n \in \cQ$ and some $\bx_n \in \cX$.
This is done by first deriving $P_{m,n+1}(q_{m,n}|x_{m,n})$ using a conditional (coordinate) sub-gradient method described by  \cite{Guoao, Nguyen,Wang12345} for all $q_{m,n} \in \cQ_m$, some $x_{m,n}\in\cX_m$,  and $m = 1,2,\ldots,M$.
Then, we use the relationship described in \eqref{qr3} to obtain the resulting $P_{n+1}(\bq_n | \bx_n )$.
Note that $P_{m,n+1}(q_{m,n}|x_{m,n})$ is the updated $P_{m,n}(q_{m,n}|x_{m,n})$.

The following theorem gives the sub-gradient of $R(s_n)$ with respect to $P_{m,n}(q_{m,n}|x_{m,n})$ for any weighted marginalized kernel at time $n$, for all $m$.
     
\begin{theorem}
 \label{lema1}
\textit{Given $\bff_{n}$,  $\bw_n$, and  an i.i.d.  sequence $ \cS,$ for any marginalized weighted kernel  defined in \eqref{normphi1},
}  
\begin{equation}\label{lem1e1}
\begin{array}{ll}
&\partial_{P_{m,n}(q_{m,n}|x_{m,n})}R(s_n)\\
& \phantom{s} = \displaystyle{ \sum_{i=1}^{n-1}\sum_{\bq_i,\bq_n\in \cQ} } \mu_{n} \alpha_{i,n}  P_i(\bq_i | \bx_{i}) \dfrac{P_n(\bq_n | \bx_{n})  }{P_{m,n}(q_{m,n}| x_{m,n})}\\
&\phantom{sdff}\times k_{\bw_{i,n}}(\bq_i,\bq_n),
\end{array}
\end{equation} 
\textit{where $\bq_i $ and $\bq_n  $ are possible quantization outputs  across the sensors at time $i$ and $n$, respectively, and
\begin{equation}\label{derivative1}
\mu_{n} = \partial_{\langle \bff_{n}, \Phi_n(\bx_n) \rangle}\ell\big(\langle \bff_{n}, \Phi_n(\bx_n) \rangle, y_n\big). 
\end{equation} 
}
\end{theorem}
\begin{IEEEproof}
Using the chain rule, we can write
\begin{equation}\label{p002e3}
\begin{array}{ll}
&\partial_{P_{m,n}(q_{m,n}|x_{m,n})}R(s_n) \\
&\phantom{ddgg}=\partial_{\langle \bff_{n}, \Phi_n(s_n) \rangle}\ell\big(\langle \bff_{n}, \Phi_n(\bx_n) \rangle,y_n\big)\\
&\phantom{tssffs}\times \partial_{P_{m,n}(q_{m,n}|x_{m,n})} \langle \bff_{n}, \Phi_n(\bx_n) \rangle. 
\end{array}\end{equation}
The first term on the right hand side (RHS) of \eqref{p002e3} is defined  as $\mu_n$ shown in  \eqref{derivative1}. The second term on the RHS of \eqref{p002e3} is
\begin{equation}\label{lem1e2}
\begin{array}{ll}
&\partial_{P_{m,n}(q_{m,n}|x_{m,n})}\langle \bff_{n}, \Phi_n(\bx_n) \rangle \\
&\phantom{s}\overset{(a)}= \partial_{P_{m,n}(q_{m,n}|x_{m,n})}\displaystyle{\sum_{i = 1}^{n-1}\sum_{\bq_i,\bq_n\in \cQ}}\alpha_{i,n}P_i(\bq_i |\bx_i)\\
&\phantom{ssssss} \times   P_n(\bq_n| \bx_n)k_{\bw_{i,n}}(\bq_i, \bq_n),
\end{array}
\end{equation}
where (\ref{lem1e2}a) holds based on \eqref{normphi1} and \eqref{weighttinstance}.
Thus, \eqref{lem1e1} is obtained by applying \eqref{qr3}  and  (\ref{lem1e2}a), where\\
$$
\partial_{P_{m,n}(q_{m,n}|x_{m,n})}P_n(\bq_n| \bx_n) = \dfrac{P_{n}(\bq_n|\bx_n) }{P_{m,n}(q_{m,n}|x_{m,n} )} .
$$
\end{IEEEproof}
   
In Theorem \ref{lema1}, we see that the size of the summation for computing the sub-gradient increases exponentially with $|\cQ|$.
To resolve this, the following proposition provides a special form of {$\partial_{P_{m,n}(q_{m,n}|x_{m,n})}R(s_n)$ for the weighted count} marginalized kernel in \eqref{wkernel3}.
    
\begin{prop}\label{prop002}
\textit{Given $\bff_{n}$,  $\bw_n$, and an i.i.d.  sequence $\cS$, $\partial_{P_{m,n}(q_{m,n}|x_{m,n})}R(s_n)$
for the kernel in \eqref{wkernel3} is written as }
\begin{align}
\partial_{P_{m,n}(q_{m,n}|x_{m,n})}R(s_n) 
&=\displaystyle{ \sum_{i=1}^{n-1} \mu_{n} \alpha_{i,n} w_{m,i}w_{m,n}P_{m,i}(q_{m,i}| x_{m,i}) } \nonumber\\
&\phantom{ssf} \times \mathds{1}(q_{m,i} = q_{m,n}, x_{m,i} = x_{m,n}), \numberthis\label{p002e2}
\end{align} 
\textit{where $\mu_{n}$ is defined in \eqref{derivative1}.
}
\end{prop}
\begin{IEEEproof}
As we  use \eqref{p002e3} in Theorem \ref{lema1}, 
we can further expand \eqref{lem1e2} using  \eqref{wkernel3} to get \eqref{p002e2}.
\end{IEEEproof}

  
Since  $P_{m,n}(q_{m,n}| x_{m,n}) $ must satisfy the summation constraint described in \eqref{mathcalp}  for all $m$ and $n$, we can update $P_{m,n}(q_{m,n}| x_{m,n})$ to $P_{m,n+1}(q_{m,n}| x_{m,n})$  through the conditional sub-gradient method with a  simplex constraint \cite{Bertsekas, Wang12345, Nguyen} 
from the summation in \eqref{mathcalp}.
We focus on \eqref{normphi1} for updating $P_n(\cdot|\cdot)$ because it will enforce quantizers across the sensors to be deterministic  after some time instance, which we will provide detailed proof of in Section \ref{Quantizaionruleol}. 

The OL setting  considered in this paper has the added complexity of dealing with data over time.
{Specifically, we update $P_{m,n}\left(q_{m,n}| x_{m,n}\right)$ at time $n$ based on the input $x_{m,n} \in \cX_m $ that is observed, as opposed to updating across all possible inputs $x_{m,n} \in \cX_m$ simultaneously as is the case in the BL setting \cite{Nguyen, Wang12345}.}
Hence, $P_{m,n+1}\left(q_{m,n+1} | x_{m,n+1}  \right)$ does  not necessarily equal the updated version of $P_{m,n}\left(q_{m,n} | x_{m,n}  \right)$, i.e., $P_{m,n+1}(q_{m,n}| x_{m,n})$, because  $x_{m,n}$ may not equal  $x_{m,n+1}$.
{The goal is to update the quantization rule $P_{m,n}(q_{m,n} | x_{m,n})$ only when the same input $x_{m,n}$ has been observed again. }

 {For this reason, we let $i_x$ denote an index for the number of times $x$ has been observed at sensor $m$ through the present time $n$. Furthermore, $n(i_x) = n$ is used to denote the time index of the most recent observation of input $x$ at sensor $m$.}
For example, if $i_x = 3$, then sensor $m$ has observed $x \in \mathcal{X}_m$ three times up through time $n$. Though $n(i_x)$ depends on $m$, we omit this in the writing of $n(i_x)$ for notational convenience.

{Thus, we must update $P_{m,n(i_x)} (q_{m,n(i_x)} | x_{m,n(i_x)})$ from time $n(i_x)$ based on \eqref{mathcalp}.}
Given $x_{n,m}$, we calculate $\mathfrak{q}_m$ from
\begin{equation}\label{pickjj}
\mathfrak{q}_m =\argmax_{q_m'\in \cQ_m} \big\| \partial_{P_{m,n(i_x)}(q_{m,n(i_x)} = q_m' | x_{m,n(i_x)})}R(s_{n(i_x)})\big\|.
\end{equation}
Then if the sub-gradient $\partial_{P_{m,n(i_x)}(\cdot|\cdot)}R(s_{n(i_x)})$ is negative, we have the following update rule inspired by \cite[Chapter 2]{Bertsekas}:
\begin{equation}\label{diffpqx1}  
\begin{array}{ll} 
&P_{m,n(i_x+1)} (q_{m,n(i_x)}| x_{m,n(i_x)})\\
&\phantom{}= (1-\eta_{n}^P)P_{m,n(i_x)}(q_{m,n(i_x)}  | x_{m,n(i_x)} )\\
&\phantom{s}-\mathds{1}(q_{m,n(i_x)} = \mathfrak{q}_m ) \eta_{n}^P\\
&\phantom{s}\times\sgn\big( \partial_{P_{m,n(i_x)}(q_{m,n(i_x)} | x_{m,n(i_x)} )}R(s_{n(i_x)}) \big),
\end{array}\end{equation}
where $\text{sgn}(\cdot)$ denotes the signum function.
Otherwise, we have 
\begin{equation}\label{diffpqx12}
\begin{array}{ll}
P_{m,n(i_x+1)} (q_{m,n(i_x)} | x_{m,n(i_x)})
= P_{m,n(i_x)} (q_{m,n(i_x)} | x_{m,n(i_x)} )
\end{array}   
\end{equation}  
for all  $q_{m,n}\in \cQ_m$
   in order to strictly satisfy the summation constraint proposed in \eqref{mathcalp} for all $m$.
   {$\eta_{n}^P \in [0,1]$ is the learning rate of the quantization rule at time $n$, which we sometimes write as $\eta_{n(i_x)}^P$ to emphasize that it would only vary based on the number of times $x$ has been observed at sensor $m$. We will see in Section~\ref{Quantizaionruleol} that $\eta_{n(i_x)}^P$ impacts the rate of convergence to a deterministic quantizer.} 

Followed by this, the quantization rule when sensor $m$ observes $x$ for the {$(i_x+1)$}th time at time $n(i_x+1)$ is 
$$
\begin{array}{ll}
&P_{m,n(i_x+1)}(q_{m, n(i_x+1)}|x_{m,n(i_x+1)})\\ &\phantom{3333kkkshow}=P_{m,n(i_x+1)}(q_{m, n(i_x)}|x_{m,n(i_x)}), 
\end{array}$$
because $x_{m,n(i_x+1)} = x_{m,n(i_x)}$.
Then,  
the expression of $P_{n+1}(\mathbf{q}_n | \mathbf{x}_n)$ follows  from \eqref{qr3}  and the fact  that
 \begin{equation}\label{quantizationsubs}
    P_{m,n+1} (q_{m,n}|x_{m,n}) = P_{m,n(i_x+1)}(q_{m, n(i_x)}|x_{m,n(i_x)} ).
\end{equation} 

\subsection{Obtaining weighted parameter vector $\bw_{n+1}$}\label{weightedparameters}
Finally, we consider obtaining the updated weight parameter vector $\bw_{n+1}$ given $\bff_n$ and $P_{n+1}(\bq_{n}|\bx_n)$.
Recall that the purpose of introducing $\bw_{n}$  is for sensor selection, i.e., to properly select the $M'$ most reliable sensors among all sensors in the classification model. 
Thus, the OL update considers the design of $\mathbf{w}_n$ and the regularization parameter $\lambda_{2,n}$ in \eqref{sop1} jointly. The update will be terminated when at most $M'$ elements of $\bw_{n+1}$ are not  $0$.  

To make a fair reliability comparison, we initialize $w_{1,1}  = \cdots = w_{M,1}$ and $\lambda_{2,1} \geq 0$.
We update $\bw_{n+1}$ using a conditional sub-gradient method to enforce a sparse solution, e.g., the projection sub-gradient method \cite{Bertsekas}.  
For a given weighted marginalized kernel, the sub-gradient of $R(s_n)$ with respect to $w_{m,n}$ can be expressed using the chain rule as
\begin{equation}\label{p001e3}
\begin{array}{ll}
\partial_{w_{m,n}}R(s_n) 
\phantom{s }=\nu_n \partial_{w_{m,n}}\langle \bff_{n}, \tilde{\Phi}_{n}(\bx_n) \rangle+ \lambda_{2,n},
\end{array}\end{equation} 
where $\tilde{\Phi}_n(\bx_n)$ defined by \eqref{phipnpuls1} is the updated marginalized mapping function at time $n$
and $\nu_n$ is given by
 \begin{equation}\label{derivative2}
\nu_n = \partial_{\langle \bff_{n},\tilde{\Phi}_n(\bx_n) \rangle}\ell\big(\langle \bff_{n}, \tilde{\Phi}_n(\bx_n) \rangle, y_n\big).
\end{equation}
Using \eqref{weighttinstance}, the gradient of the decision function is obtained as
\begin{equation}\label{p001e5}
\begin{array}{ll}
 \partial_{w_{m,n}}\langle \bff_{n},\tilde{\Phi}_n(\bx_n) \rangle 
& = \partial_{w_{m,n}}\displaystyle{\sum_{i = 1}^{n-1}\sum_{\bq_i,\bq_n\in \cQ}}\alpha_{i,n}P_{n+1}(\bq_n|\bx_n)\\
&\phantom{lll}\times P_i(\bq_i|\bx_i) k_{\bw_{i,n}}(\bq_i,\bq_n).
\end{array}\end{equation} 

As in Section \ref{updatequantizationrules}, the computational requirement for solving \eqref{p001e5} increases with $|\cQ|$.
The following proposition gives an efficient expression for $\partial_{w_{m,n}}R(s_n)$ based on the kernel in \eqref{wkernel3}.



\begin{prop}\label{prop001}
\textit{Given $\bff_{n}$,  $P_{n+1}(\bq_n | \bx_n)$,  and an i.i.d.  sequence $\cS$, $\partial_{w_{m,n}}R(s_n)$ for the kernel in
\eqref{wkernel3} is}
\begin{equation}\label{p001e2}
\begin{array}{ll}
&\partial_{w_{m,n}}R(s_n)\\
& \phantom{s} = \lambda_{2,n} +\displaystyle{ \sum_{i=1}^{n-1}\sum_{q_{m,i}\in \cQ_m}\Big( \nu_n \alpha_{i,n}  w_{m,i}}P_{m,i}(q_{m,i} | x_{m,i})\\
&\phantom{sh} \times    P_{m,n+1}(q_{m,n} | x_{m,i})\mathds{1}(q_{m,i} = q_{m,n}, x_{m,i} = x_{m,n})\Big).
\end{array}
\end{equation}
\end{prop}

\begin{IEEEproof}
We start by writing \eqref{p001e3} in terms of \eqref{derivative2} and \eqref{p001e5}. Using \eqref{wkernel3}, we obtain the following expression for \eqref{p001e5}:
\begin{equation}\label{p001e7} 
\begin{array}{ll}
\eqref{p001e5}
&=\displaystyle{ \sum_{i=1}^{n-1}\sum_{q_{m,i} \in \cQ_m} \alpha_{i,n}  w_{m,i} } P_{m,i}(q_{m,i} | x_{m,i}) 
\\
&\phantom{sh} \times P_{m,n+1}(q_{m} | x_{m,n}) \mathds{1}(q_{m,i} = q_{m,n}, x_{m,i} = x_{m,n}).
\end{array}
\end{equation}
Thus, we  obtain \eqref{p001e2} using \eqref{wkernel3}, \eqref{p001e3}, and \eqref{p001e7}.
\end{IEEEproof}

With this in hand, we can employ a projected sub-gradient method to update $\bw_{n+1}$ while satisfying the  constraint in \eqref{weightedparameterconstraint}.
Given the objective of selecting $M' \leq M$ sensors to be kept on, the update is terminated once we reach a time instance $n$ such that $M - M'$ elements of $\mathbf{w}_{n+1}$ are zero. 
Specifically, at time $n$, to get a sparse solution, we update
\begin{equation}\label{bw1}
    w_{m,n+1} = \begin{cases}
    \dfrac{\widetilde{w}_{m}M}{\| \widetilde{\bw} \|_1}, &\text{if }\|\mathbf{w}_n\|_0 <M- M',\\
    w_{m,n},&\text{if }\|\mathbf{w}_n\|_0 =M- M',
    \end{cases}
\end{equation}
to maintain constraint \eqref{weightedparameterconstraint}, where $\|\bw_n \|_0$ is the count of the number of
non-zero elements, and the $m$th element of $\widetilde{\bw}$ is
\begin{equation}\label{bw12}
\widetilde{w}_{m} =\max\{0, w_{m,n} - \eta_{n}^{\bw}\partial_{w_{m,n}}R(s_n)\},
 \end{equation}
where $\eta_{n}^{\bw}>0$ is learning rate with respect to $\bw_n$.
 { $\eta_n^{\mathbf{w}}$ controls the training time prior to disabling $M-M'$ sensors. As the value of $\eta_n^{\mathbf{w}}$ is increased, this time will be shortened.}
To ensure that exactly $M'$ sensors are enabled, we can design the regularization constant  $\lambda_{2,n}$ such that at most one sensor is set to zero in \eqref{bw1} at each time $n$:
\begin{equation}\label{lambda2n}
\begin{array}{ll}  
\lambda_{2,n}
&=\min\big\{\{w_{m,n}-\partial_{w_{m,n}}\langle \bff_{n}, \tilde{\Phi}_n(\bx_n) \rangle \\  
&\phantom{kk}\times \partial_{\langle \bff_{n}, \tilde{\Phi}_n(\bx_n) \rangle}\ell\big(\langle \bff_{n}, \tilde{\Phi}_n(\bx_n) \rangle,y_n\big)\}_{m=1}^{M}\big \}.
\end{array}\end{equation} 
There are also other potential settings for $\lambda_{2,n}$  which will result in $\|\mathbf{w}_n\|_0 = M - M'$ while yielding possibly different values of $\mathbf{w}_{n+1}$.
Other conditional sub-gradient methods can be used as long as the sparse solution is enforced.


Algorithm \ref{a1} summarizes the entire MSOKSQ procedure developed throughout this section.
{In practice, this algorithm will continue running indefinitely, with the value of $n$ incrementing each time a new collection of measurements is generated across the sensors.}
{For the weighted count marginalized kernel \eqref{wkernel3}, Propositions 1 and 2 indicate that the quantization rule and weight parameter updates can be done independently at time $n$ across each sensor $m$.}

Note that in the case of the soft margin loss function from \eqref{hingeloss},  we can express \eqref{alpha1234}, \eqref{derivative1}, and \eqref{derivative2} based on
\begin{equation}\label{derivative}
\begin{array}{ll}
\partial_{\langle \bff_{n}, ~\Phi_{n}(\bx_{n}) \rangle}\ell\big(\cdot, \cdot\big)
=\begin{cases}
-y_{n}, &\text{if } y_{n} \langle \bff_{n}, \Phi_{n}(\bx_{n}) \rangle \leq \rho,\\
0, &\text{otherwise}.
\end{cases}
\end{array}
\end{equation}
 {We will employ this non-differentiable loss function form in our theoretical analysis in Section \ref{CA} and numerical experiments in Section \ref{NPA}.}

\section{Theoretical Analysis}\label{CA}
In this section, we theoretically analyze our MSOKSQ algorithm.
We first provide assumptions for the OL and BL approaches similar to \cite{Kivinen, Nguyen, Wang12345} (Section \ref{assumptions}).
Then, we study the relationship between the OL and BL approaches (Section \ref{cbob}). 
Followed by this, we will characterize the update behavior of the quantization rules and sensor weights from MSOKSQ in the OL setting (Section \ref{Quantizaionruleol}).
Finally, we conduct a convergence analysis that compares the OL and BL approaches (Section \ref{cb1}).    

\subsection{Definitions and Assumptions}\label{assumptions}
Consider a fixed i.i.d. sequence of measurements $\mathcal{S}$. We assume there is a $\Psi>0$ such that the marginalized mapping functions of the BL and OL settings in \eqref{mgk} and \eqref{mmftime} satisfy $\|\Phi(\bx_n)\|,~\|\Phi_{n}(\bx_n) \| \leq \Psi$, for all $n$ and $s_n \in \cS$. 
Additionally, we assume that $\bff$ derived by the BL approach is bounded such that $\| \bff \|\leq F$ for any $\bff \in \cH$, where  
\begin{equation}\label{UF}
  F\geq \dfrac{c \Psi}{\lambda_1} \text{ and } \lambda_1 >0, \text{ for some } c >0.
\end{equation}  
The value of $c>0$ is derived based on the loss function $\ell(\cdot, \cdot)$ which we assume satisfies the Lipschitz condition \cite{Kivinen}: 
\begin{equation}\label{lipfn}    \|\ell(\xi_1, y) - \ell(\xi_2, y)\|\leq c \|\xi_1-\xi_2\|,
\end{equation}
where  $\xi_1,~\xi_2 \in \mathbb{R}$ and $y \in \mathcal{Y}$.
Hence, for the OL approach, given $\|\bff_1\| = 0$, we bound $\|f_n \| \leq \frac{c \Psi}{\lambda_1}$ for $n = 1,2,\ldots, N$, which follows from
    $\|\bff_{n+1}\|
       \leq (1-\lambda_1 \eta_n)\|\bff_n\|+ \eta_n c\Psi$
based on \eqref{diffbff1}
and \eqref{dirrf} for all possible $\bff_n$ and $n$.
As we set $\|\Phi_{n}(\bx_n)\| \leq \Psi$ and  $\|\bff_n \| \leq \frac{c \Psi}{\lambda_1}$, we obtain
    $\|\partial_{\bff_n} R(s_n) \| \leq 2c\Psi$ from \eqref{diffbff1}. 

{The learning rate for $\mathbf{f}_n$ can be chosen to be $\eta_n = \eta_1n^{-0.5}$ where $\eta_1\in (0, 1]$, which is a standard setting found in OL problems, e.g., \cite{Guoao, Koppel, Koppel1, Kivinen} due to its theoretical guarantees.
This setting is desirable in our problem given the convergence bound relationships it will produce with prior batch learning training approaches, as we will see in Section \ref{cb1}.}

 {Finally, we define a set $\Gamma$ that contains (i) all possible solutions $( \gamma_n)_{n =1 }^{N}$ for the OL setting, as well as (ii) all possible solutions $\gamma$ for the BL setting.}
Moreover, we consider a set $\Gamma_{d}\subset \Gamma$ for the deterministic quantizer such that the weights of the decision function $\mathbf{f}_n$ and the weighted parameter vector $\mathbf{w}_n$ are jointly derived based on the rules defined in \eqref{determinsquantizationrules}, for both the OL and the BL setting.   

\subsection{Connecting OL  and BL Approaches}\label{cbob}
We quantitatively compare the behavior of our OL methodology to previously studied BL approaches \cite{Nguyen, Wang12345}. Following prior work on OL \cite{Kivinen, Koppel, Koppel1, Scholkopf}, we will compare the average instantaneous regularized risk  function for the OL setting, i.e.,
\begin{equation}\label{airf}
 R_{\text{avg}}(\cS) = \frac{1}{N}\sum_{n=1}^{N}R(s_n),
\end{equation}
  with $R_{\text{reg}}(\cS)$ for the BL setting defined in \eqref{sop1}.     

Before we make the connection, we first define
 \begin{equation}\label{rmrk1}
R_{\bff}(s_n) = \ell\big( \langle {\bff}, \Phi_{n}(\bx_n) \rangle,y_{n}\big) +\dfrac{\lambda_1}{2} \|{\bff}\|^2 + {\lambda_{2,n}}\|\bw_n\|_1
\end{equation}
to be the OL risk function defined in \eqref{rmrk} for a constant weight on decision function $\mathbf{f}$ \cite{Kivinen}, i.e., the BL case. Due to the convexity of $R(\cdot)$, we know that $
\left \langle \partial_{\bff_n} R(s_n), \bff_n -  \bff\right\rangle \geq
R_{\bff}(s_n) - R(s_n)
$. Thus, the inequality
   \begin{equation}\label{relatoin1}
  \begin{array}{ll}
& \frac{1}{\eta_n}\left(\| \bff_n -\bff \|^2\right) -\frac{1}{\eta_{n+1}}\| \bff_{n+1} -\bff\|^2  \\
 & \phantom{s1111111}\overset{\phantom{(b)}}\geq-4\eta_nc^2\Psi^2+2R(s_n)-2R_{\bff}(s_n)\\
 &\phantom{ssf11111111d}+\left(\frac{1}{\eta_n} -\frac{1}{\eta_{n+1}}\right)\|\bff_{n+1} - \bff \|^2 
\end{array} 
  \end{equation}
from \cite{Kivinen} holds for our setting. With this inequality, we obtain the following lemma:



\begin{lemma}\label{lemma1}
 \textit{Given  an i.i.d.  sequence $\mathcal{S}$ of $N$ time samples, suppose Algorithm  \ref{a1} is executed with $\eta_n = \eta_{1} n^{-0.5}$ and $\eta_{1}\in (0,1]$. For any solution set sequence $(\gamma_n)_{n=1}^{N} \in \Gamma_n$, we have } 
 \begin{equation}\label{le31}
\begin{array}{ll}
&R_{\text{avg}}(\cS)= \displaystyle{\sum_{n = 1}^{N}}\frac{cF\left\| \Phi_{n}(\bx_n)-  \Phi(\bx_n)\right\|}{2N} \\
&+\displaystyle{\sum_{n = 1}^{N}}\frac{c\big\|\lambda_{2,n} \|\bw_n\|_1 - \lambda_2\|{\bw} \|_1\big\|}{2N}+ R_{\text{reg}}\left(\cS \right) + O(N^{-0.5})
\end{array}    
\end{equation} 
\textit{where $O(\cdot)$ is Big $O$ notation, for some $c >0$}.
\end{lemma}
\begin{IEEEproof}
We first show the sequence of inequalities in (\ref{le1}).
  \begin{figure*}
   \begin{equation}\label{le1}
   \begin{array}{ll}
\dfrac{\| \bff_1 - {\bff} \|^2}{\eta_1} - \dfrac{\| \bff_{n+1} - {\bff} \|^2}{\eta_{n+1}}
&\overset{(a)}= \displaystyle{\sum_{n = 1}^{N}}\bigg[ \dfrac{\| \bff_n - {\bff} \|^2}{\eta_n} -\dfrac{\| \bff_{n+1} - {\bff} \|^2}{\eta_{n+1}}+\left(\dfrac{1}{\eta_{n}}-  \dfrac{1}{\eta_{n+1} }\right) \| \bff_{n+1} - {\bff}\|^2 \bigg] \\
& \overset{(b)}\geq-8c^2\eta_1 \Psi^2 N^{0.5}+2\displaystyle{\sum_{n=1}^N} R(s_n)- 2\displaystyle{\sum_{n=1}^N}R_{\bff}(s_n)-\dfrac{4F^2 N^{0.5}}{\eta_1}\\
 & \overset{(c)}\geq \dfrac{-8c^2 \eta_1\Psi^2}{ N^{0.5}}+ \dfrac{1}{N}\displaystyle{\sum_{n=1}^N}2R(s_n)-2R_{\text{reg}}(\cS)  -\dfrac{4F^2 }{N^{0.5}\eta_1}\\    
&\phantom{show me }-c \displaystyle{\sum_{n = 1}^{N}}\dfrac{\Big(F \left\| \Phi_{n}(\bx_n)-  \Phi(\bx_n)\right\| + \big\| \lambda_{2,n}\|\bw_n\|_1 - \lambda_2\|{\bw} \|_1\big\|\Big)}{2N}.
\end{array}   
   \end{equation}
   \hrulefill
\end{figure*}
Inspired by \cite{Kivinen}, we obtain (\ref{le1}a) by  adding
 $\begin{array}{ll}
 \sum_{n = 1}^{N}\frac{\| \bff_{n+1} - {\bff} \|^2}{\eta_{n}}-\sum_{n = 1}^{N}\frac{\| \bff_{n+1} - {\bff} \|^2}{\eta_{n}}.\end{array} 
 $
 The first three terms on the RHS of (\ref{le1}b) are derived using  $\sum_{n=1}^{N}\eta_n \leq 2 \eta N^{0.5}$ as shown in \cite{Kivinen} and   \eqref{relatoin1}.  
  The fourth term holds because $\| \bff_{n+1} -  {\bff} \|^2 \leq 4F^2$ from the triangle inequality,
 $  \sum_{n = 1}^{N}  \frac{1}{\eta_{n}}- \frac{1}{\eta_{n+1} }  = \frac{1}{\eta_1} - \frac{1}{\eta_{N+1}} = \frac{1}{\eta_1} - \frac{(N+1)^{0.5}}{\eta_1}, $
  and $-N^{0.5} \leq 1-(N+1)^{0.5}$.
Given this, we derive (\ref{le1}c) using
   \begin{align} 
 &\dfrac{1}{N}\displaystyle{\Big( \sum_{n=1}^{N}}R_{\bff}(s_n)\Big)-R_{\text{reg}}(\cS)   \nonumber\\
&\phantom{sh}\overset{{(a)} }\leq \dfrac{1}{N}\displaystyle{\sum_{n=1}^{N}} \Big\|\ell\left(\langle \bff, \Phi_n(\bx_n) \rangle,y_n\right)  \nonumber\\
&\phantom{shdsj}- \ell\left(\langle {\bff}, \Phi(\bx_n) \rangle,y_n\right)+ (\lambda_{2,n}\|\bw_n\|_1 - \lambda_2\|\bw \|_1) \Big\|  \nonumber\\ 
&\phantom{sh}\overset{{(b)}}{ \leq}\dfrac{c}{N}\displaystyle{\sum_{n=1}^{N}}\Big(F\| \Phi_{n}(\bx_n) -\Phi(\bx_n)\|  \nonumber\\  
&\phantom{show}\ + \big\|\lambda_{2,n} \|\bw_n\|_1 - \lambda_2\|{\bw}\|_1   \big\|\Big),\numberthis \label{le4}
 \end{align}
where (\ref{le4}a) follows from Jensen's inequality and the fact that  the weights of  $\langle \bff, \Phi_n(\bx_n) \rangle$ and $\langle \bff, \Phi(\bx_n) \rangle$ are the same, and
(\ref{le4}b) follows from \eqref{lipfn} and the Cauchy Schwartz inequality. 
Finally, note that
\begin{equation}\label{le2}
\dfrac{\| \bff_1 - \bff \|^2}{\eta_1} - \dfrac{\| \bff_{n+1} - \bff \|^2}{\eta_{n+1}} \leq \frac{F^2}{\eta_1} - 0
\end{equation}
 based on the facts  $\bff_1 = \mathbf{0}$ and  $\| \bff_{n+1} - \bff \|\geq 0$.  
Then, \eqref{le31} is justified considering \eqref{le1} and \eqref{le2}.
\end{IEEEproof}
   

Lemma \ref{lemma1} establishes a relationship between the risk function of the OL and BL approaches: there is a gap between $R_{\text{avg}}(\cS)$ and $R_{\text{reg}}(\cS)$
when there is a larger variance in the marginalized mapping function $\Phi_n(\cdot)$ and sensor selection weights $\mathbf{w}_n$ over time.
We next explore how our update rules in Algorithm 1 can further refine the bound in \eqref{le31}.



\subsection{Quantization Rules and Weight  Parameter Vector Analysis}\label{Quantizaionruleol}
In \eqref{le31}, the first term is dependent on
 the quantization rules $P_n(\cdot|\cdot)$ and the weighed parameter vector $\bw_n$ while the second term  relates to the weight parameter vector.
We are thus interested to study how Algorithm \ref{a1} affects the updates of the quantization rules and the weighted parameter vector over time. In our analysis, we will assume the soft margin loss function \eqref{hingeloss} for the binary classification problem.



The quantization rules across the sensors  have significant impacts on the classification performance, as we will show numerically in Section \ref{NPA}. Prior works in decentralized signal detection \cite{Nguyen, Guoao, Wang12345} have not considered the effect of quantization rules.
Thus, we are interested in studying the update behaviors of the quantization rules  based on the weighted count marginalized kernel in \eqref{wkernel3}.
Before we provide our Proposition \ref{propbl}, some facts need  to be addressed. 


 
Recall $n(i_x) \in \{ 1,2,\ldots, N\}$ is the time index at which sensor $m$ observes the $i_x$-th instance of $x$.  
The values of $\alpha_{i,n(i_x)}$ and $\mu_{1,{n(i_x)}}$ in  \eqref{alpha1234} and \eqref{derivative1} are affected by  $\rho$  defined in \eqref{hingeloss}.
For any  $x\in \cX$, considering \eqref{alpha1234}, \eqref{derivative1}, \eqref{p002e2}, and \eqref{derivative}, we have  that
\begin{equation}\label{propqble4}
    \text{sgn}\big(\alpha_{i,n(i_x)}\mathds{1}(x_{m,i} =  x_{m,n(i_x)})\big) =-\text{sgn}(\mu_{1,n(i_x)}).
\end{equation}
This indicates from \eqref{p002e2} that 
\begin{equation}\label{propqble5}
\partial_{P_{m,n(i_x)}(q_{m,n(i_x)} |x_{m,n(i_x)})}R(s_{n(i_x)}) <0,   \forall q_{m,n(i_x)}\in \cQ_m.
\end{equation} 
Hence, the update of $P_{m,n(i_x+1)}(q_{m,n(i_x)}  | x_{m,n(i_x)})$ occurs when $\partial_{P_{m,n(i_x)}(q_{m,n(i_x)}  |x_{m,n(i_x)})}R(s_{n(i_x)})< 0$, for some $q_{m,n(i_x)}$.
Specifically,  $P_{m,n(i_x)}(q_{m,n(i_x)}= \mathfrak{q}_m|x_{m,n(i_x)})$ is increased by $\eta_{n(i_x)}^P -\eta_{n(i_x)}^P P_{m,n(i_x)}(q_{m,n(i_x)} = \mathfrak{q}_m|x_{m,n(i_x)}) > 0$  while $P_{m,n(i_x)}(q_{m,n(i_x)} = q_m|x_{m,n(i_x)})$, i.e., for all other $q_m \neq \mathfrak{q}_m$ is reduced by $\eta_{n(i_x)}^P$ based on \eqref{diffpqx1}.  $\mathfrak{q}_m$ is defined in \eqref{pickjj}.   
As we expect  $P_{m,n}(q_{m,n}|x_{m,n})$ for a particular value of $x$ to be unimodal in practice, we  initialize  
\begin{equation}\label{propqble42}
P_{m,n(i_x) }(q_{m,n(i_x) } |x_{m,n(i_x) })= \max_{q_m \in \cQ_m} \{P_m(q_m|x)\},  
\end{equation} 
for $n(i_x) \in \{1,2,\ldots,N \}$, $m = 1,2,\ldots,M$,
and $i_x = 1$ for all $x\in \cX_m$.
\begin{prop}\label{propbl} 
\textit{
Given an i.i.d. sequence  $\cS$, suppose the quantization rules are initialized based on \eqref{propqble42},
and $\ell(\cdot, \cdot)$ is defined in \eqref{hingeloss} with a properly selected $\rho$ such that 
$\{\partial_{P_{m,n(i_x)}(q_{m,n(i_x)} |x_{m,n(i_x)})}R(s_{n(i_x)})<0\}_{i_x =1}^{N_x}$. Then, the updates of $P_{m,n(N_x)}(q_{m,n(N_x)} |x_{m,n(N_x)})$    based on Algorithm \ref{a1} and the kernel in \eqref{wkernel3} can be described as}
\begin{equation}\label{propqble1}
    P_{m,n(N_x+1)}(q_{m,n_{N_x}}|x_{m,n_{N_x}})=\begin{cases} 
    &1 - O\big((1-\eta)^{N_x}\big),\\
    &\phantom{ssss  ss}\text{if }q_{m,n_{N_x}} = q_m,\\
    &O\big((1-\eta)^{N_x}\big),\\
    &\phantom{ssss  ss}\text{otherwise,}
    \end{cases}
\end{equation}    
 \textit{for a fixed  $q_m$ defined in \eqref{propqble42} and 
 \begin{equation}\label{propqble2}
  \eta = 1- \prod_{i_x = 1}^{N_x}(1-\eta^P_{n(i_x)})^{\frac{1}{N_x}}.
 \end{equation}
 }
\end{prop}
     
\begin{IEEEproof}
This proof starts with generalizing the expression of $P_{m,n(N_x+1)}(q_{m,n(N_x)} \neq  q_m|x_{m,n(N_x)})$, which is given by
\begin{equation}
\begin{array}{ll}
  &P_{m,n(N_x+1)}(q_{m,n(N_x+1)} \neq q_m |x_{m,n(N_x)} )\\
  &\phantom{ssss ss}= \displaystyle{\prod_{i_x = 1}^{N_x}}(1-\eta^P_{n(1)})P_{m,n(1)}(q_{m,n(1)}  \neq q_m |x_{m,n(1)})\\
  &\phantom{sss}\\
  &\phantom{ssss ss}=(1-\eta)^{N_x}P_{m,n(1)}(q_{m,n(1)}  \neq q_m|x_{m,n(1)})\\ 
  &\phantom{sss}\\
  &\phantom{ssss ss} = O\big((1-\eta)^{N_x}\big)
  \end{array}  
\end{equation}  
using \eqref{diffpqx1}, \eqref{pickjj}, \eqref{propqble4}-\eqref{propqble42}, and \eqref{propqble2}.
Followed by this, we use  \eqref{mathcalp} to obtain 
$P_{m,n(N_x+1)}(q_{m,n(N_x)} =  q_m |x_{m,n(N_x)}) = 1-O\big((1-\eta)^{N_x}\big).$
\end{IEEEproof}
Based on Proposition \ref{propbl}, if we set the learning rate $\eta_{n(i_x)}^P = 1$ for all $n(i_x)$, the quantizer of sensor $m$ will become a deterministic quantizer  with respect to observation $x$ once $N_x > 1$.
{In this case, the quantization rules of the sensors become the same, and the sensor selection strategy is effectively disabled, since it relies on quantizer diversity.
More generally, if $\eta_{n(i_x)}^P \in (0,1)$ for all $n(i_x)$ and $m = 1,2,\ldots,M$, the quantizer approaches deterministic behavior asymptotically as the number of observations $N_x$ increases, with larger $\eta_{n(i_x)}^P$ producing a faster rate of convergence. Our numerical experiments in Section VI will confirm that convergence is reached even for small values of $\eta_{n(i_x)}^P$.}

Since our primary goal is to train the decision function and the quantization rules across the sensors, we are also interested in the case when the sensor selection strategy is not used (i.e., $M' = M$).
Additionally, the algorithms proposed by \cite{Nguyen} and \cite{Wang12345} are the same when  $\lambda_2 = 0$ and $\bw = \mathbf{1}$ is fixed.
Hence, we can compare $R_{\text{avg}}(\cS)$ from our OL setting with $R_{\text{reg}}(\cS)$ defined in \cite{Nguyen} when $\lambda_2 = \lambda_{2,n} = 0$ and $\bw = \bw_{n}$, which indicates that
 \begin{equation}\label{assumption234}
k_{\bw_{\cdot, n}}(\cdot,\bq_n) = k_{\bw}(\cdot,\bq_n), \text{for } n=1,2,\ldots,N,
\end{equation}
where $k_{\bw_{\cdot,n}}(\cdot,\cdot)$ and $k_{\bw}(\cdot,\cdot)$ correspond to the OL and BL settings, respectively.
Based on this, we next aim to analyze the performance of our OL methodology against that of the BL setting assuming the BL optimal solution drawn from $\Gamma_d$. 

\subsection{Convergence  Analysis}\label{cb1}
We will now leverage Proposition 3 to tighten our result in Lemma 1. In doing so, we will assume the sensor selection strategy is not employed, which makes our OL setting directly comparable to prior work \cite{Nguyen} for the BL approach as discussed above. In these works,
the relationship between the true risk function and the empirical risk function of the BL approach (i.e., the first term on the RHS of \eqref{sop1}) have been shown to follow
 \begin{equation}\label{errorbound}
 \begin{array}{ll}
&\displaystyle{\argmin_{\gamma_{\text{true}}\in \Gamma}} ~  \mathbb{E}_{\bX,Y}\left[\mathbb{E}_{\bQ} \left[\ell\left(f(\bQ),y\right)\right | \bX,Y ]\right]\\
&\phantom{show me}=\displaystyle{\argmin_{\gamma \in \Gamma_{d}}\sum_{n=1}^{N}}\frac{\ell(\langle \bff, \Phi(\bx_n) \rangle ,y_{n})) }{N}+ O(N^{-0.5}),
\end{array}
\end{equation}  
where  $\gamma_{\text{true}}$ is the optimal solution of the true risk function assuming  the fusion center and all sensors know $P(\bX,Y)$. 
 $\gamma \in \Gamma_d$ indicates that 
 \begin{equation}\label{determinsquantizationrules2}
     P(\bq_n|\bx_n) = \begin{cases}
 1, &\text{ if } \bq_n = {\bq },\\
      0, &\text{ otherwise, }
     \end{cases} \text{ ~for some }\bq \in \cQ.
 \end{equation}  


        The following theorem shows that a similar relationship exists between $R_{\text{avg}}$ from the OL approach and $R_{\text{reg}}$ from the BL approach.
\begin{theorem}\label{proconnection}
 \textit{Consider the soft margin loss $\ell(\cdot, \cdot)$ in \eqref{hingeloss} with a properly selected $\rho$ as in Proposition \ref{propbl}, $\eta_n = \eta_{1} n^{-0.5}$, and an i.i.d.  sequence $\mathcal{S}$ 
 such that all sensors have observed all possible $x \in \cX_m$ at least $N_x+1 $ times at time $N'<n$. Furthermore, suppose $\bw_n = \bw$, and $\lambda_{2,n} = \lambda_{2} = 0$, for all $n$.
If the  quantization rules of the OL approach are initialized as in \eqref{propqble42}, and the deterministic quantization rules of the BL approach described in \eqref{determinsquantizationrules} follow \eqref{propqble42},
then, for large $N$,}
 \begin{equation}\label{le312}
R_{\text{avg}}(\cS)=  R_{\text{reg}}\left(\cS \right) + O(h(N)),
\end{equation}
\textit{where}  
\begin{equation}
h(N) = \max \left\{N^{-0.5}, O((1-\eta)^{N_x}) \right\},    
\end{equation}   
\textit{and $\eta$ is defined by Proposition \ref{propbl}.}
\end{theorem}   
\begin{IEEEproof}  
    Given the fact that $\lambda_{2,n} = \lambda_2 = 0$ and $\bw_n = \bw$ for all $n$, \eqref{le31} can be re-written based on Lemma \ref{lemma1} as
        \begin{equation}\label{le3121}
R_{\text{avg}}(\cS)= \displaystyle{\sum_{n = 1}^{N}}\frac{cF\left\| \Phi_{n}(\bx_n)-  \Phi(\bx_n)\right\|}{2N} + R_{\text{reg}}\left(\cS \right) + O(N^{-0.5}).
    \end{equation}
     The norm in \eqref{le3121} can be written as
    \begin{equation}\label{le31331}
     \begin{array}{ll}
     &   \| \Phi_{n}(\bx_n)-  \Phi(\bx_n)\|\\
        &\phantom{sdf}\overset{\phantom{(a)}}=\Big\|\displaystyle{\sum_{\bq_n\in\cQ}}k_{\bw_{\cdot, n}}(\cdot,\bq_n)\big(P_{n}(\bq_n | \bx_n)- P(\bq_n|\bx_n)\big)\Big\|,
     \end{array}
    \end{equation}
where (\ref{le31331}) follows \eqref{assumption234}. 
Note that $P_{n}(\bq_n|\bx_n)$ and  $P(\bq_n|\bx_n)$ are quantization rules across the sensors with respect to the OL and BL approaches, respectively.

  At time $n = N'$,  each sensor has observed all possible $x$ at least $N_x$ times.
  Thus, once $n \geq N'$,
    we can approximate the quantization rules across the sensors based on Proposition \ref{propbl} and  \eqref{mathcalp} as
    \begin{equation}
    P_{n}(\bq_n|\bx_n)= \begin{cases}
    1-O\big((1-\eta)^{N_x }\big), & \text{ if } \bq_n =  {\bq},\\
    O\big((1-\eta)^{N_x }\big), & \text{ otherwise, } 
    \end{cases}
    \end{equation}
    for a fixed $\bq \in \cQ$.
Now, writing
    \begin{equation}\label{le3133}
    \begin{array}{ll}  
&\displaystyle{\sum_{n = 1}^{N}}\frac{cF\left\| \Phi_{n}(\bx_n)-  \Phi(\bx_n)\right\|}{2N} = \displaystyle{\sum_{n = 1}^{N' }} \frac{cF\left\| \Phi_{n}(\bx_n)-  \Phi(\bx_n)\right\|}{2N}\\
&\phantom{ssss ss ss sssssssssss}+ \displaystyle{\sum_{n = N'+1}^N}    \frac{cF\left\| \Phi_{n}(\bx_n)-  \Phi(\bx_n)\right\|}{2N},
\end{array}
    \end{equation}
  we  can   approximate
     \begin{equation}\label{convergencebound1}
  \displaystyle{\sum_{n = 1}^{N'}}    \frac{cF\left\| \Phi_{n}(\bx_n)-  \Phi(\bx_n)\right\|}{2N} \leq  \frac{N' cF\Psi}{2N} =O\left (\dfrac{1}{N} \right) 
    \end{equation}
   by the triangle inequality and for large $N$,  and
    \begin{equation}\label{convergencebound2}
    \begin{array}{ll}
  \displaystyle{\sum_{n = {N'+1}}^N}    \frac{cF\left\| \Phi_{n}(\bx_n)-  \Phi(\bx_n)\right\|}{2N}
  = O((1-\eta)^{N_x })
    \end{array}\end{equation}
    using \eqref{le31331} and the fact that 
   the initialized quantization rules of the OL and the deterministic quantization rules of the BL follow \eqref{propqble42}.
\eqref{le312} then follows from \eqref{convergencebound1} and \eqref{convergencebound2}.
\end{IEEEproof}
  
  Note that in the case when  $\eta = \eta_1^P= \eta_2^P = \cdots = \eta_N^P = 1$, \eqref{le312} can be written based on Proposition \ref{proconnection} as 
\begin{equation}\label{newapprox}
R_{\text{avg}}(\cS) = R_{\text{reg}}(\cS)+ O(N^{-0.5}).
\end{equation} 
 Comparing Theorem \ref{proconnection} and Lemma \ref{lemma1}, the performance of our OL methodology converges to that of BL setting \cite{Kivinen} assuming the BL optimal solution belonging to $\Gamma_d$ is considered.

\begin{remark}\label{remark1}
Placing a deterministic quantizer at each sensor indicates that  the size of $|\cX|$ would be reduced to $|\cQ|$.
In this case, for any $\bx \in \cX$, the class label of $\bx$ would always the same as that of $\bq$, where $\bq$ is the corresponding quantization output of $\bx$.
Thus, we are also interested in analyzing the classification performance between a $\text{log}_2|\cQ_m|-$bit deterministic quantizer and an infinite-resolution deterministic quantizer. 
In this paper, each sensor using an infinite-resolution deterministic quantizer would imply that the quantization operation can be neglected, i.e., $\cQ_m  = \cX_m$ and that the   sensors are equally reliable. In this particular setting, Algorithm \ref{a1} can be seen as the native online regularized-risk minimization algorithm (NORMA) \cite{Kivinen}.
NORMA described by \cite{Kivinen} trains the weights   of the decision function that can be  implemented in the proposed multi-sensor model.
Given the fact that the classification performance of  NORMA \cite[Theorem 4]{Kivinen} is approximately equal to the BL setting when  infinite-resolution deterministic quantizers are considered, we use NORMA as the an upper bound on achievable performance of MSOKSQ in Section \ref{NPA}. \end{remark}

 \section{Numerical Performance Analysis}\label{NPA}
We will now conduct a numerical evaluation of our MSOKSQ methodology. We will consider two datasets for classification tasks distributed across sensors: a 2-class randomly generated dataset (Section \ref{2b}), and the 3-class Iris dataset from the University of California, Irvine (UCI) open source repository \cite{UCI} (Section \ref{2c}). Before presenting these results, we will discuss our evaluation setup (Section \ref{2a}).
\subsection{Simulation Setup}\label{2a}
In each experiment,
we set the learning rates for $\bff_n$ to $\eta_{n} = 0.1n^{-.5}$ (satisfying the conditions in Lemma 1 and Theorem 2), $P_{n}(\cdot|\cdot)$ to $\eta_{n}^{P} = 0.1$,  and $\bw_n$ to $\eta_{n}^{\bw} = 0.5$. 
{For regularization parameters, we set $\lambda_1 = 0.1$ (which satisfies the constraint from \eqref{alpha1234} for each $\eta_{n(i_x)}^P$), and $\lambda_{2,n}$ according to \eqref{lambda2n}. We find that these settings lead to stable training performance. Additionally, we use a margin parameter $\rho = 1$ (hinge loss).  }
We initialize $\bff_1 = \mathbf{0}$, $\bw_1 = \mathbf{1}$, and $P_{n}(\cdot|\cdot)$ based on \eqref{propqble42}.
The simulations are done using the weighted count marginalized kernel in \eqref{wkernel3}.

Inspired by \cite{Gersho}, we define the quantization points as uniform partitions over the range $[-A, A]$, i.e.,
\begin{equation}\label{quantizationpt}
\cQ_m = \left\{\frac{(2d +1 -|\cQ_m|)A}{|\cQ_m|}\right\}_{d = 0}^{|\cQ_m|-1}.
\end{equation}

To evaluate the performance of MSOKSQ, we consider several baselines/variants of Algorithm \ref{a1}:
\begin{itemize}
\item \textit{NORMA}: Introduced by \cite{Kivinen}, NORMA can be used for the proposed multi-sensor mode, where each sensor forwards its observation directly to the fusion center. 
The detailed discussion of this baseline can be found in Remark \ref{remark1}.

\item \textit{MSOKSQ without updating $P_{m,n}$ and $w_n$}: This can be considered as NORMA because $P_{m,n}(\cdot|\cdot)$ and $\bw_n$ are not being updated by Algorithm \ref{a1} for all $n$.

\item \textit{MSOKSQ without updating $\bw_n$}: Here, Algorithm \ref{a1} is executed assuming $\bw_n=\mathbf{1}$ for all $n$. When $M=1$, this corresponds to the results in \cite{Guoao}.

\item \textit{MSOKSQ with $\log_2[|\cQ_m ]$-bit deterministic quantizer}: In this case, Algorithm \ref{a1} is executed assuming $\bw_n=\mathbf{1}$ for all $n$.

\item \textit{MSOKSQ optimally selects $M'$}: Here, Algorithm \ref{a1} is performed given $M'$ sensors need to be enabled.

\item \textit{MSOKSQ randomly selects $M'$}: Here, $M'$ sensors are randomly enabled, which can be done by reassigning the value of $w_{m,n}$ produced by Algorithm 1 randomly. 
\end{itemize}
\begin{figure}[!t]
\centering
\includegraphics[width=3.5in]{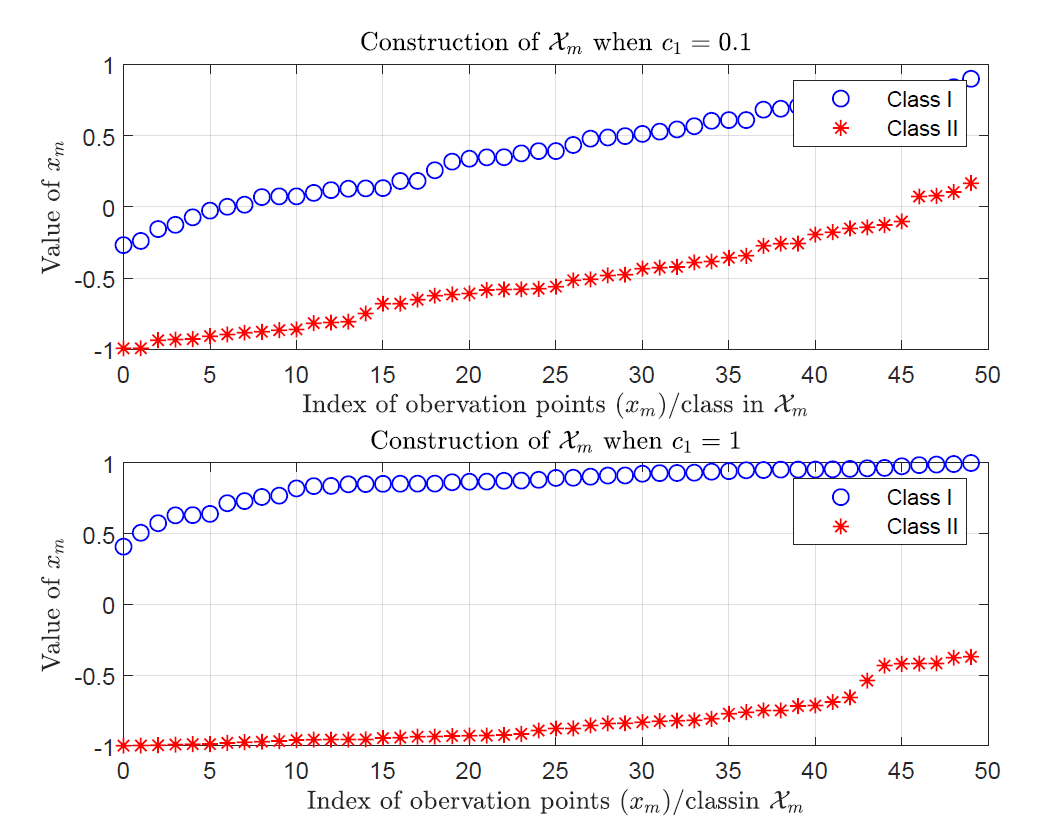}
\caption{Example of generated $\cX_m$ for a given $m$ based on \eqref{gendata1} with $c_1 = 0.1$ and $c_1= 1$.}
\label{figdatagen}
\end{figure} 
 
\begin{figure}[!t]
\centering
\includegraphics[width=3.5in]{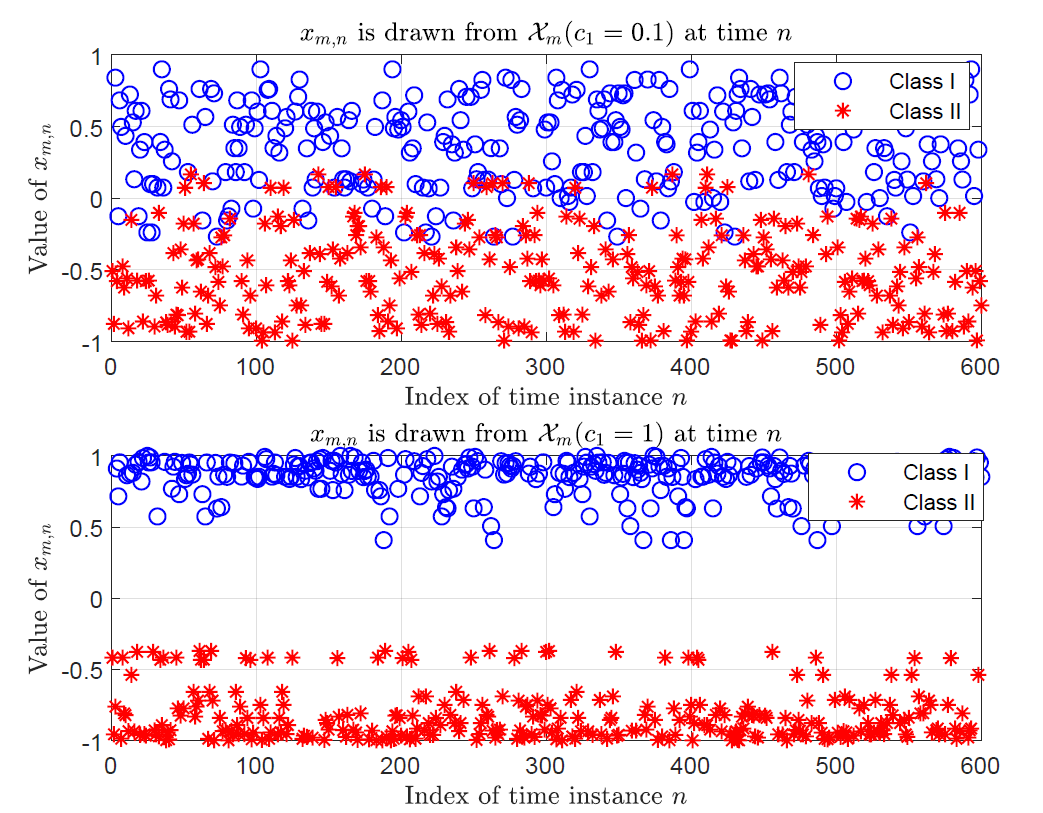}
\caption{Example of generated samples $x_{m,n}$ over time $n = 1,...,600$ from the non-separable ($c_1=0.1$) and separable ($c_1 = 1$) $\mathcal{X}_m$ from Fig. \ref{figdatagen}.}
\label{figdatagen1}
\end{figure} 
  
\subsection{Binary Classification  Problem from Synthetic Data}\label{2b}

We first consider a binary classification task, i.e., $y \in \{-1, 1\}$. In our evaluation, we aim to emulate different degrees of (i) measurement variation across sensors and (ii) separability of the dataset between classes. To do so, we generate elements $x_m$ of  $\mathcal{X}_m$ for sensor $m$ according to
\begin{equation}\label{gendata1}
    x_{m} = \begin{cases}
    y- 0.75\left((1+c_1)c_2+c_2^2\right), & \text{ when } y = 1, \\
    y+ 0.75\left((1+c_1)c_2+c_2^2\right), & \text{ when } y = -1,
    \end{cases}
\end{equation}
where $c_1 \in (0,1]$ is a constant and $c_2 \sim \mbox{Uni}(0,1)$ is sampled from a uniform distribution over $(0,1)$ for each measurement. In this way, each $x_m \in [-1,1]$, and the observation sets $\cX_1, \cX_2, \ldots, \cX_M$ are each unique. At time $n$, observation $x_{m,n}$ is assumed to be randomly selected from $\mathcal{X}_m$.

Note that we can vary the degree to which the classes are separated through proper selection of $c_1$.  
For example, at a single sensor $m$,  Fig. \ref{figdatagen} shows the construction of  $\cX_m$ with 100 data points, where 50 points are generated for each class, for $c_1 = 0.1$ and $c_1 = 1.0$.
Fig. \ref{figdatagen1} shows samples generated over time $n = 1,...,600$ for each of the $\mathcal{X}_m$ in Fig. \ref{figdatagen}.
We see that $\cX_m$ with $c_1 = 0.1$ is a linearly  separable dataset while $\cX_m$ with $c_1 = 1$ is non-separable.


In the following simulations, we
consider $N=600$ time instances.    
We assume that $P_{m,n}(q_{m,n}|x_{m,n})$ is initialized based on \eqref{mathcalp} for all possible $q_{m,n} \in \cQ$ and $x_{m,n} \in \cX_m$.
By default, each sensor is equipped with a $3$-bit uniform quantizer, where $\cQ_m$ is defined in \eqref{quantizationpt}.
The observation set $\mathcal{X}_m$ for sensor $m$ is generated according to (70) with $|\cX_m|=20$ and  $10$ data points/class. 
The fusion center first classifies the quantization outputs $\bq_n$ forwarded by the sensors using
\begin{equation}\label{decisionfunction123321}
\langle \bff_n, k_{\bw_{\cdot,n}}(\cdot,\bq_n) \rangle = \sum_{i = 1}^{n-1}\alpha_{i,n}\sum_{\bq_i \in \cQ}P_{i}(\bq_i|\bx_i)k_{\bw_{i,n}}(\bq_i,\bq_n),
\end{equation}
where $k_{\bw_{i,n}}(\bq_i,\bq_n)$ follows  the weighted count kernel in  \eqref{wkernel2}.
Then, the fusion center updates the  solution $\gamma_n$ based on Section \ref{OLA}.
The accuracy classification rate (ACR) is used as our classification performance metric, which is
\begin{equation}\label{ACR1}
    \text{ACR}(n) = \displaystyle{\sum_{j=1}^n} \frac{1}{n}\mathds{1}\left(\text{sgn}\left(\langle \bff_j, k_{\bw_{\cdot,j}}(\cdot,\bq_j) \rangle\right) =y_j\right),
\end{equation} 
for $n = 1,...,N$. {This metric quantifies the signal classification accuracy of the WSN over time, which allows us to assess the performance of our online learning methodology and compare the impact of different components of our algorithm (i.e., the quantization, sensor selection, and decision function updates). Similar metrics (e.g., error rate, total loss) have been employed in prior works for decentralized OL \cite{Kivinen, Koppel1}.}

\begin{figure}[!t]
\centering
\includegraphics[width=3.5in]{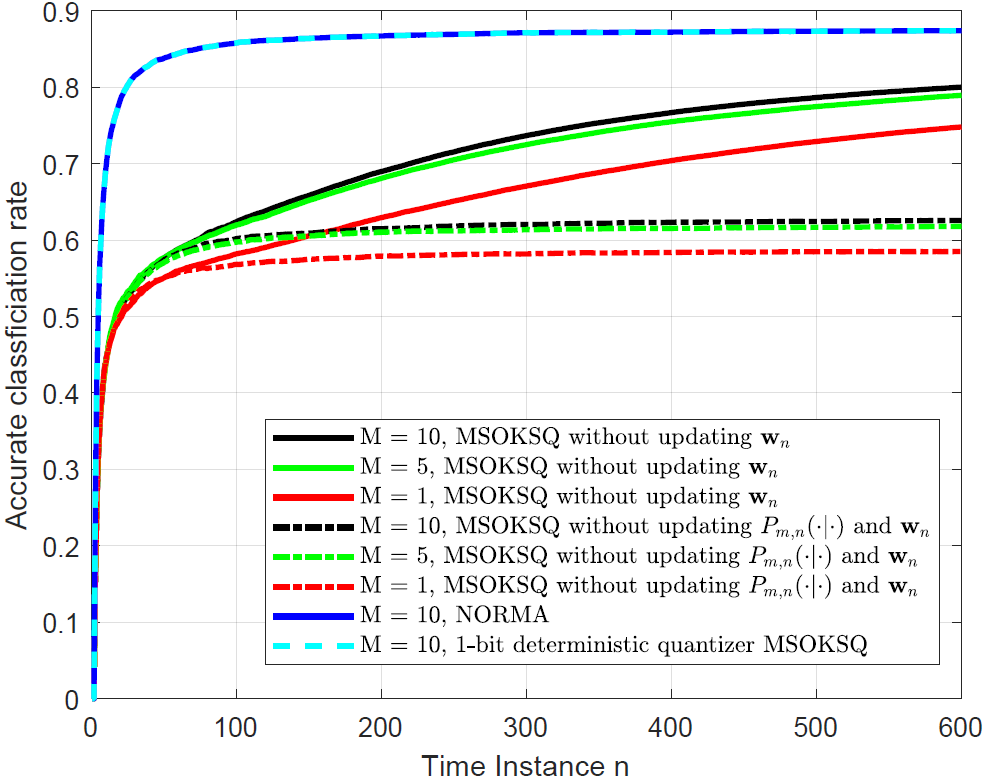}
\caption{ACR performance for Algorithm \ref{a1} between (i) MSOKSQ without updating $\bw_n$ and (ii) MSOKSQ without updating $\bw_n$ and $P_n(\cdot|\cdot)$  for varying sensors $M$, assuming a 3-bit quantizer is used for each sensor.  
 The performance of NORMA and MSOKSQ when each sensor uses a 1-bit deterministic quantizer are also given for $M= 10$.}
\label{okSQransensor2}
\end{figure}

In Fig. \ref{okSQransensor2}, we study how the number of sensors and optimization of the  quantization rules affect the ACR performance.
We consider the number of sensor $m = 1,5,$ and $10$.
We assume that   $c_1 =0.1$ in \eqref{gendata1}, i.e., the dataset is non-separable.
We see that ACR for both MSOKSQ without updating  $\bw_n $ and MSOKSQ without updating $P_{m,n}(\cdot,\cdot)$ and $\bw_n$ increase with the number of sensors.
As the time instance increases, MSOKSQ without updating $\bw_n$ begins to offer substantial improvements in ACR performance over MSOKSQ without updating $P_{m,n}(\cdot|\cdot)$ and $\bw_n$ (up to ~20\% by $n = 600$), which demonstrates the importance of optimizing the quantization rules.
Moreover, the ACR performance  of MSOKSQ without updating $\bw_n$ approaches the ACR performance of NORMA in the limit because the randomized quantizers become  deterministic quantizers (Proposition \ref{propbl}). 
Further, when each sensor
uses  a 1-bit deterministic quantizer, the ACR performance of  MSOKSQ overlaps that of NORMA, as we discuss in Remark 2 below.

\begin{figure}[!t]
\centering
\includegraphics[width=3.5in]{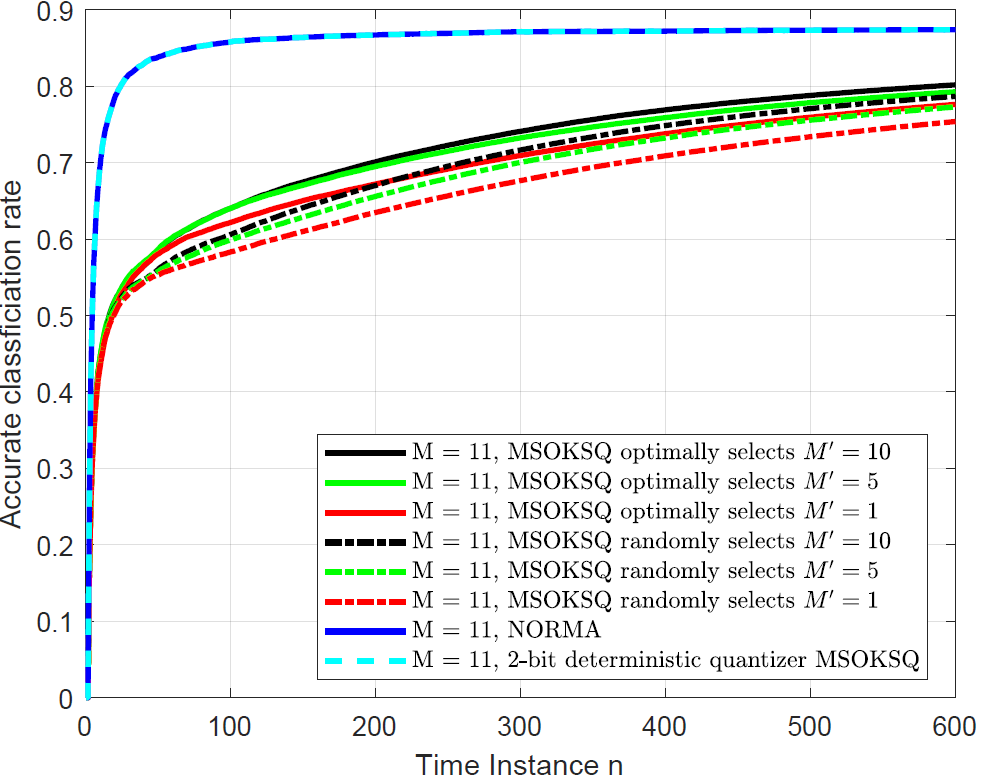}
\caption{ACR performance of Algorithm \ref{a1} between  MSOKSQ with (i) optimized and (ii) random sensor selection strategies  when
 $M'=1, 5,$ and $10$ sensors are enabled among $M = 11$ sensors, and a $3$-bit quantizer is used for each sensor. 
 The  performance of NORMA and MSOKSQ when each sensor uses a 2-bit deterministic quantizer are also shown for  $M = 11$. } 
\label{okSQransensor1}
\end{figure}

  \begin{remark}
Sensors are equally reliable when  deterministic quantizers are considered. 
As Fig. \ref{okSQransensor2}, Fig. \ref{okSQransensor1}, and Fig. \ref{okSQirissensor1}$-$ Fig. \ref{okSQirissensor3} show,  if we properly select the precision of the deterministic quantizers, i.e., $|\cQ_m|$, for all $m$, the ACR performance of the $\text{log}_2[|\cQ_m|]$-bit deterministic quantizer  achieves the same performance as NORMA.
\end{remark}

Fig. \ref{okSQransensor1} analyzes the ACR performance  for both optimal and random sensors selections when the total number of sensor is $M=11$.
We include three different cases of $M' = 1, 5,$ and $10$  sensors  enabled, given sensors equip $3$-bit quantizers.
We see that the ACR performance of MSOKSQ increases over time for both optimal and random sensor selection strategies, with optimal sensor  selection outperforming random sensor selection for all choices of $M'$. Also, while the performance decreases when $M'$ decreases, we see that our optimization strategy provides ACR robustness as the available sensor selection budget changes from $M' = 5$ to $10$.
Additionally, we see that the ACR performance  of MSOKSQ optimal/random sensor  selections approach NORMA as the time instance increases for all $M'$.
As in Fig. \ref{okSQransensor2}, this is due to the fact that the randomized quantizers become  deterministic in the limit.
Finally, when each sensor uses a 2-bit deterministic quantizer, we see that the ACR performance of  MSOKSQ overlaps NORMA, consistent with Remark 2.

\begin{figure}[!t]
\centering
\includegraphics[width=3.5in]{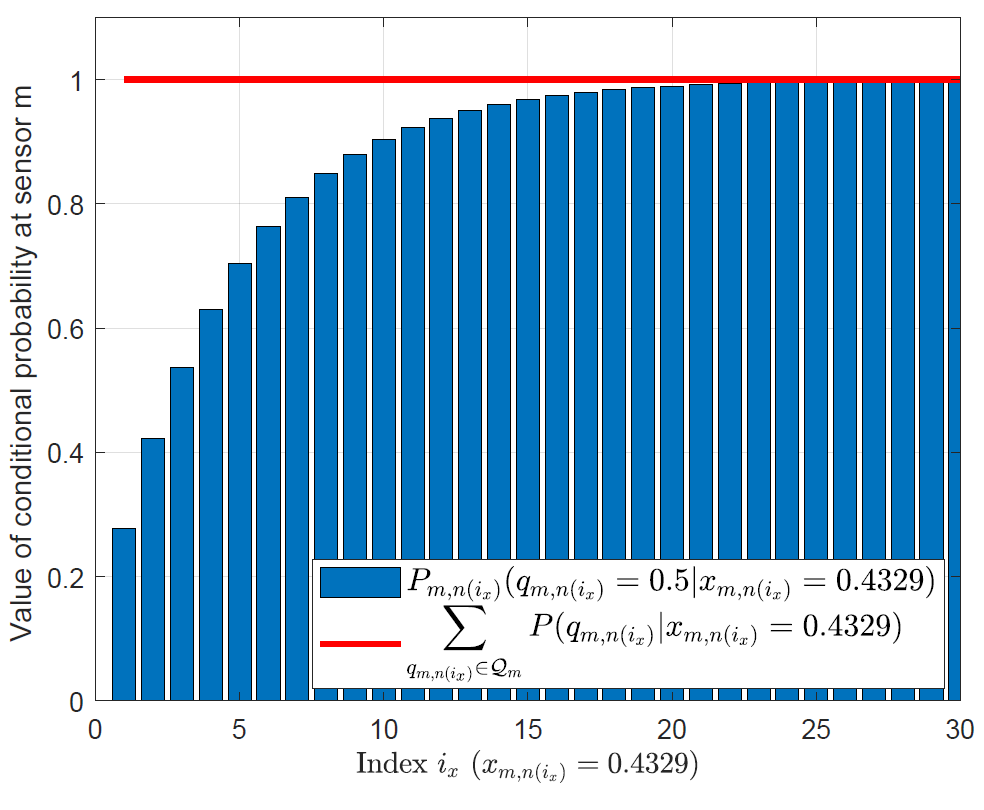}
\caption{Demonstrating the  value of $P_{m,n(i_x)}(q_{m,n(i_x)} =0.5|x_{m,n(i_x)}=0.4329)$ for $i_{x} = 1,2,\ldots,30$, where $i_x$ is the $i_x$-th time that sensor $m$ observes $0.4329$.}
\label{conditional}
\end{figure}

\begin{figure}[!t]
\centering
\includegraphics[width=3.5in]{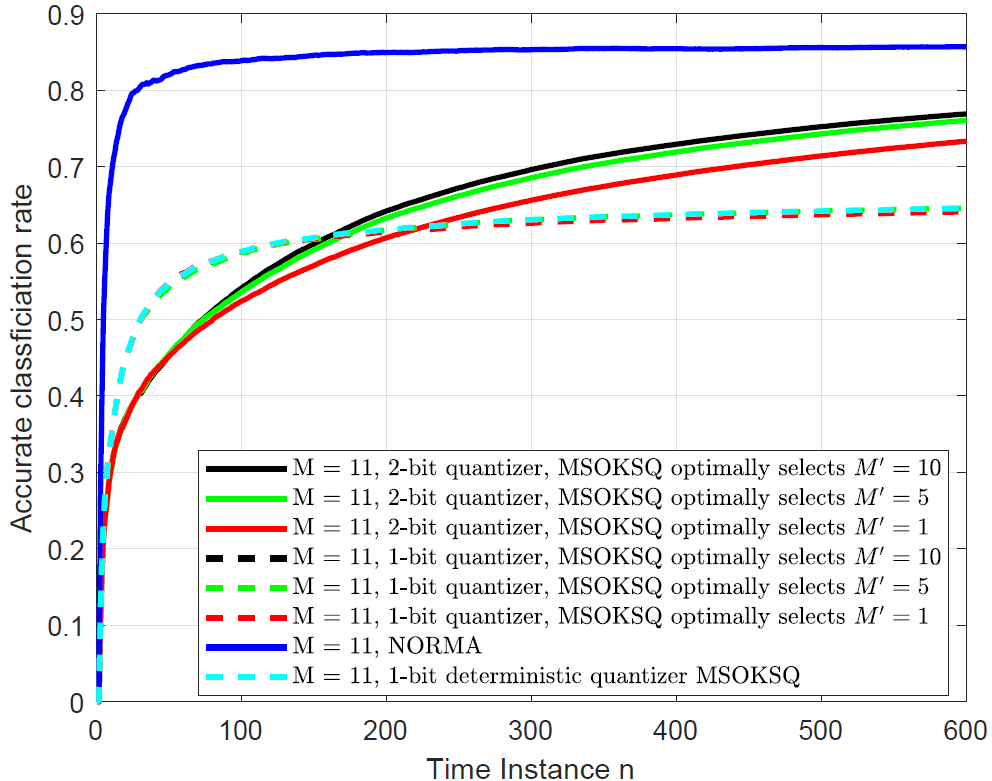}
\caption{The ACR performance for Algorithm \ref{a1} between  MSOKSQ with (i) 1-bit and (ii) 2-bit quantizers for  $M'=1, 5,$ and $10$  sensors to be selected out of a total of $M = 11$. 
 The performance of NORMA and MSOKSQ when each sensor uses a 1-bit deterministic quantizer is also given for $M = 11$.}
\label{okSQirissensor1}
\end{figure}

   Fig. \ref{conditional} demonstrates the quantization rule behavior of MSOKSQ at a particular value of $x \in \mathcal{X}_m$ ($x = 0.4329$).
  The settings in of Fig. \ref{conditional} are the same as Fig. \ref{okSQransensor1}.
Given $\cQ_m = \{-0.5, 0.5\}$  defined in \eqref{quantizationpt}, the quantization rule of sensor $m$ is initialized such that
\begin{equation}\label{iniprh12345}
\begin{array}{ll}
 &P_{m,n(1)(q_{m,n(1)}}=0.5|x_{m,n(1)} = 0.4329)\\
 &\phantom{s}= \displaystyle{\max_{q_{m,n(1)} \in \cQ_m}} P_{m,n(1)}(q_{m,n(1)} |x_{m,n(1)}=0.4329).
 \end{array} 
 \end{equation}
 Fig. \ref{conditional} indicates that $P_{m,n(i_x)}(q_{m,n(i_x)}=0.5|x_{m,n(i_x)} = 0.4329)$ approaches 1 with $i_x$ (the number of times $x$ has been observed).
 Thus, Fig. \ref{conditional} is consistent with Proposition  \ref{propbl}.




\subsection{Multi-Class Classification Problem on Iris Dataset}\label{2c}
Although our presentation throughout the paper has focused binary classification, Algorithm \ref{a1} can handle multi-class classification as well.
In the 3-class Iris dataset  \cite{UCI}, each Iris plant belongs to one of three classes $\cY \in \{ [-1, \phantom{-}1], [1, -1], [1, \phantom{-}1]\}$,  and thus $\mathbf{y}_n = [y_{1,n}, y_{2,n}]$ for each sample $n$ in our setting. We assume that the observation set $\mathcal{X}_m$ for each sensor $m$ is the petal width of the Iris plants such that $\cX_1 = \cX_2=\cdots=\cX_M$.
If we assume each sensor equips a 2-bit quantizer, then the quantization points are $\mathcal{Q}_m = \{-3,-1,1,3\}$ \cite{Gersho}.
Over $N = 600$ time instances, each sensor makes its observations $x_{m,n}$ that are assumed to be drawn from $\cX_m$ randomly per time instance, for all $m$. 

 To address multi-class within our methodology, we will employ the standard One Versus All (OVA) technique, where a decision function is trained for each label dimension separately \cite{Scholkopf, Bishop,Rifkin}.
At each time instance $n$, then, the update in Algorithm 1 is run once for each decision function independently. In the Iris setting, we need two decision functions, where the first and the second decision functions classify the first label, $y_{1,n}$, and the second label, $y_{2,n}$, respectively. 
The ACR for the multi-class problem is defined in \eqref{ACR2}, where $\bff_{1,n}$ and $\bff_{2,n}$  are the weights of the decision functions that will be use to determine $\mathbf{y}_n$, for $n = 1,2,\ldots,N$.
\begin{figure*}[!t]
\begin{equation}\label{ACR2}
       \text{ACR}(n) = \displaystyle{\sum_{j = 1}^n}\frac{ \mathds{1}\left(\mathrm{sgn}\left(\langle \bff_{1,j},    k_{\bw_{\cdot,j}}(\cdot,\bq_j) \rangle\right) =y_{1,j} , \mathrm{sgn}\left(\langle \bff_{2,j},     k_{\bw_{\cdot,j}}(\cdot,\bq_j) \rangle\right) =y_{2,j}\right)}{n},
\end{equation}
\hrulefill
\end{figure*}
Fig. \ref{okSQirissensor1}  demonstrates how the number of quantization bits affects the ACR for the multi-class problem.
We assume the total number of sensor is $M = 11$.
We compare the cases when each sensor produces 1-bit and 2-bit outputs assuming $M' = 1,5,$ and $10$   sensors are enabled.
Initially, the ACR performance of the 1-bit quantizer exceeds that of the 2-bit quantizer because the quantization rules of the 1-bit quantizer become  deterministic faster than   the 2-bit quantizer.
However, a $1$-bit quantizer maps $x_{m,n}$ onto one of the only quantization points.
Hence, as the time instance increases, the ACR performance of using a 2-bit quantizer/sensor begins to outperform a 1-bit quantizer/sensor as expected.
Intuitively, the system requires at least two bits to reach maximum accuracy in a 3-class problem.
 Additionally, we see that enabling more sensors (i.e., increasing $M'$) only has a noticeable impact in the 2-bit case, which indicates that coupling the quantization design across sensors is more impactful when the sensors have higher precision. Finally, when each sensor uses a 1-bit deterministic quantizer, the ACR performance   overlaps  optimal sensor selection with a 1-bit quantizer.

\begin{figure}[!t]
\centering
\includegraphics[width=3.5in]{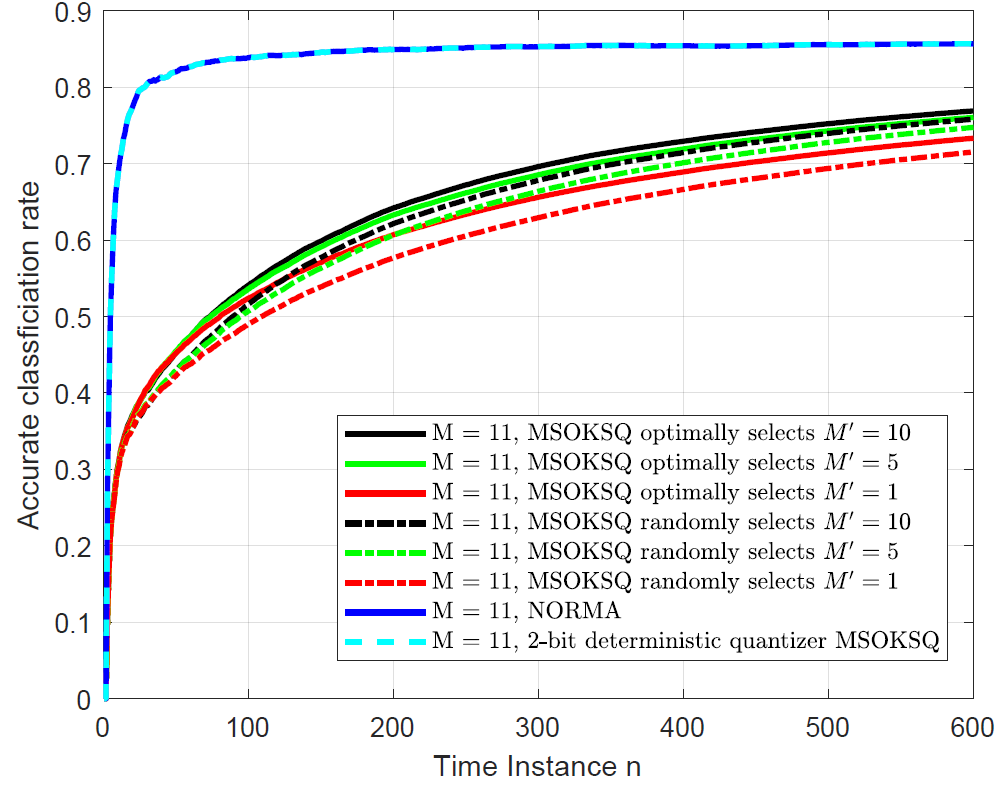}
\caption{The ACR performance for Algorithm \ref{a1} between MSOKSQ with (i) random and (ii) optimal sensor selection strategies.  $M' = 1,5$, and $10$ sensors are assumed to be enabled, and a 2-bit quantizer is used for each sensor.
The  performance  between NORMA and MSOKSQ when each sensor uses a 2-bit deterministic quantizer when $M = 11$.
}
\label{okSQirissensor2}
\end{figure}

{Fig. 8 presents a similar experiment to Fig. 5, showing the ACR performance of both optimal and random sensor selection when $M=11$.}
We assume each sensor is equipped with a $2$-bit quantizer.
We compare three different cases when $M' = 1, 5,$ and $10$ number of sensors are assumed to be enabled.
Fig. \ref{okSQirissensor2} yields similar conclusions to Fig. \ref{okSQransensor1}, i.e., with optimal strategies outperforming random selection. 
  Moreover, when each sensor uses  a 2-bit deterministic quantizer, we see that the ACR performance of MSOKSQ overlaps NORMA, consistent with Remark 2. 

\begin{figure}[!t]
\centering
\includegraphics[width=3.5in]{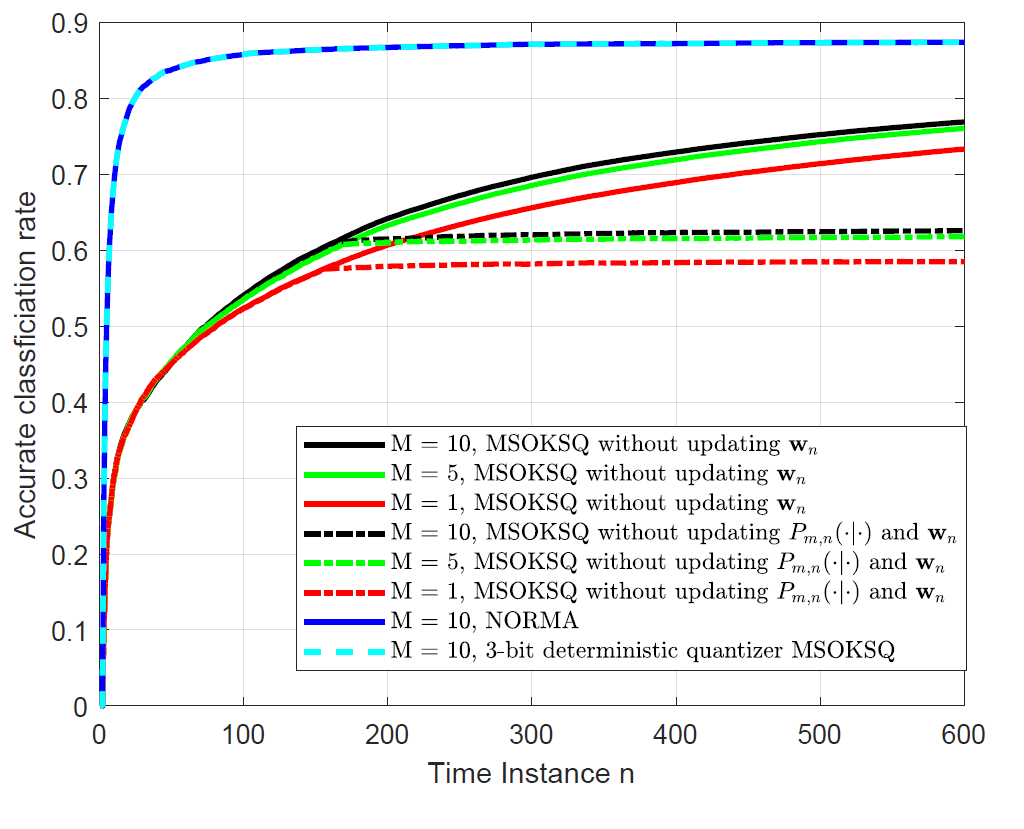}
\caption{ACR performance for Algorithm \ref{a1} between  MSOKSQ with optimal and random sensor selection.   $M'=1, 5,$ and $10$   sensors are assumed to be enabled and  a $2-$bit quantizer is used for each sensor. 
The  performance of  NORMA and MSOKSQ when each sensor uses a 3-bit deterministic quantizer is also shown for $M = 11$.}
\label{okSQirissensor3}
\end{figure}

Fig. \ref{okSQirissensor3} presents a similar experiment to Fig. \ref{okSQransensor2}, studying how the number of sensors and optimal quantization rules affect the ACR performance defined in \eqref{ACR2}.
We consider the number of sensors to be $M = 1, 5,$ and $10$.
Fig. \ref{okSQirissensor3} shows similar trends to Fig. \ref{okSQransensor2}, i.e., the quantization rule updates provide significant benefits over time.  
 Finally, when each sensor uses a 3-bit deterministic quantizer, the ACR performance of  MSOKSQ overlaps NORMA,  consistent with Remark 2.


\section{Conclusion and Future Work}
  In this paper, we developed MSOKSQ, a novel methodology for signal classification in wireless sensor networks that jointly optimizes quantization rules, sensor selection, and decision weights for online learning.
  Our theoretical analysis revealed the behavior of the quantization rule updates based on the proposed algorithm, and established the convergence bound relationship with prior batch learning training approaches.
 Finally, we provided a numerical evaluation on two classification tasks which demonstrated the improvements in learning accuracy obtained by different components of MSOKSQ.
   
In our future work, we are interested in explicating the role the proposed system model in various applications in wireless communications.
 In particular, we aim to leverage MSOKSQ to learn wireless system parameters such as transmit beamformers, quantizer designs, and error control coding. 
 {Future work can also investigate online signal classification in wireless sensor networks for settings where no fusion center exists \cite{Lin9562522,Li9426904}, e.g., through cooperative consensus formation of locally trained decision weights, sensor selection, and quantization rules (or decision rules) at each sensor. A key challenge here will be accounting for potential processing resource limitations across the sensors.}

 \begin{IEEEbiography}[{\includegraphics[width=1in,height=1.25in,clip,keepaspectratio]{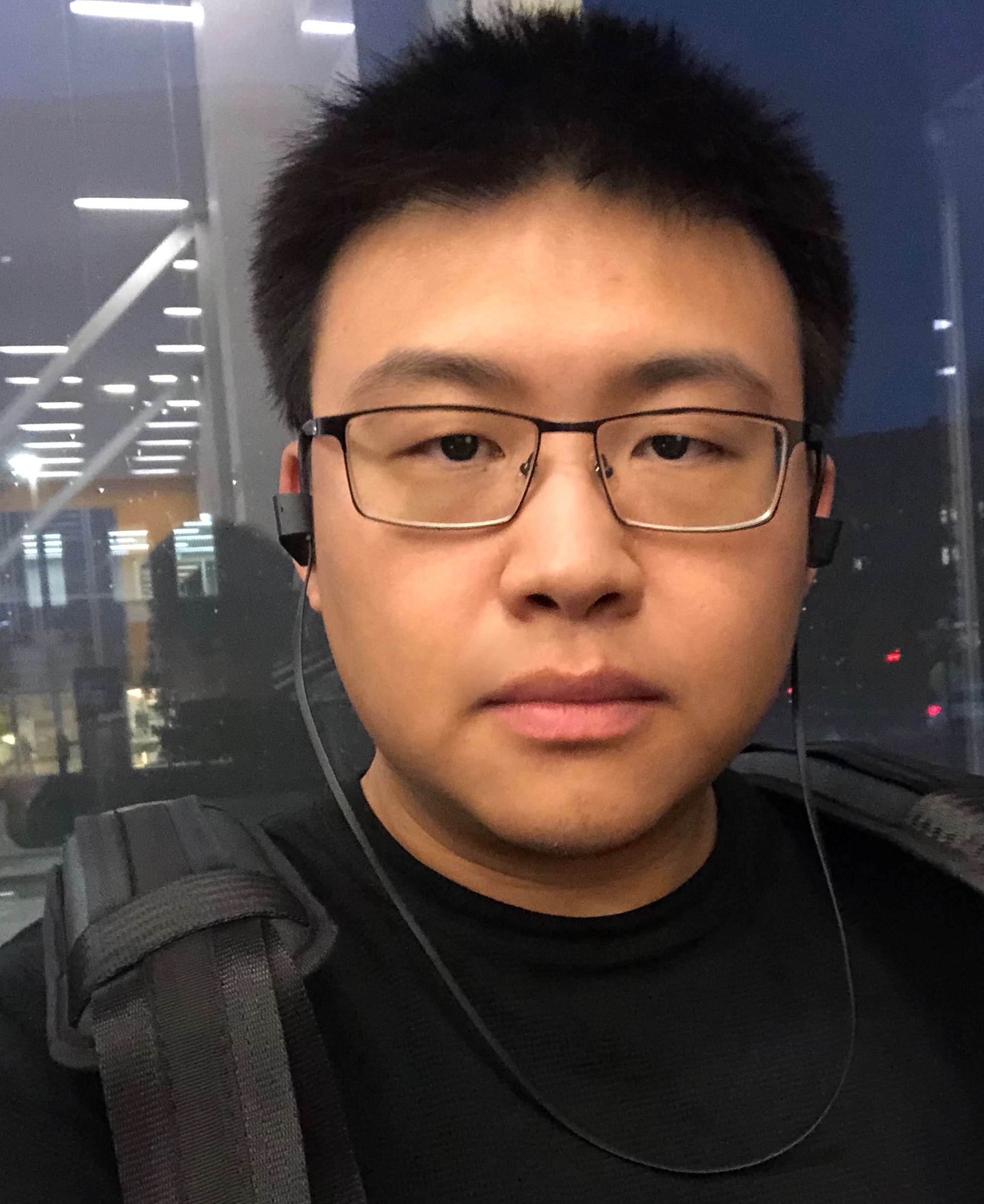}}]{Jing Guo} received the B.S. degrees in electrical engineering and mathematics from University of Maryland, College Park, MD, USA, in 2014. He received the Ph.D degree in electrical engineering from Purdue University, West Lafayette, IN, USA, in 2021. In summer 2014, he was a lab assistant at NASA, Greenbelt, MD, USA, for the outer-space measurement project. In summer 2016, 2017, and 2018, he was a research assistant with MIT Lincoln lab, Lexington, MA, USA, where his prime focuses are signal processing and optimal rate allocation in wireless communication.
 In summer 2019 and 2020, he was a research assistant with Naval Research Laboratory, Washington DC, where he focused on applying machine learning in wireless sensor network. He joined NXP Semiconductor in summer 2021. His research interests are sub-6GHz and beyond-5G wireless systems, multiple-input multiple-output communications, MIMO array processing, and machine learning.

\end{IEEEbiography}

\begin{IEEEbiography}[{\includegraphics[width=1in,height=1.25in,clip,keepaspectratio]{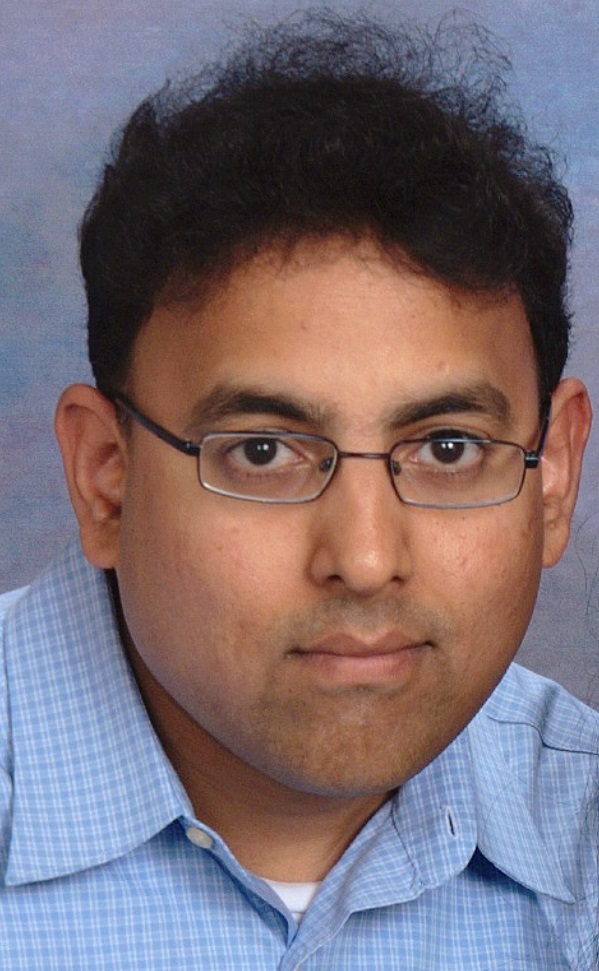}}]{Raghu G. Raj}
 (Senior Member, IEEE) received the Ph.D. degree in Electrical Engineering from The University of Texas at Austin in 2007; and his undergraduate degrees in Computer Science and Electrical Engineering from Washington University in St. Louis. He is currently a Senior Research Scientist and the Head of the Radar Imaging and Target ID Section, Radar Division, of U.S. Naval Research Laboratory (NRL), Washington DC, where he leads the research and development of advanced methods in radar imaging, detection, and target identification with applications to various U.S. DoD funded programs. He has over 70 publications in various international journals, conferences, and technical reports. His research interests span various signal/image processing, machine learning, and inverse problems in radar and remote sensing. He holds eight U.S. patents and is a recipient of the NRL Alan Berman Publication Award.
\end{IEEEbiography}

\begin{IEEEbiography}[{\includegraphics[width=1in,height=1.25in,clip,keepaspectratio]{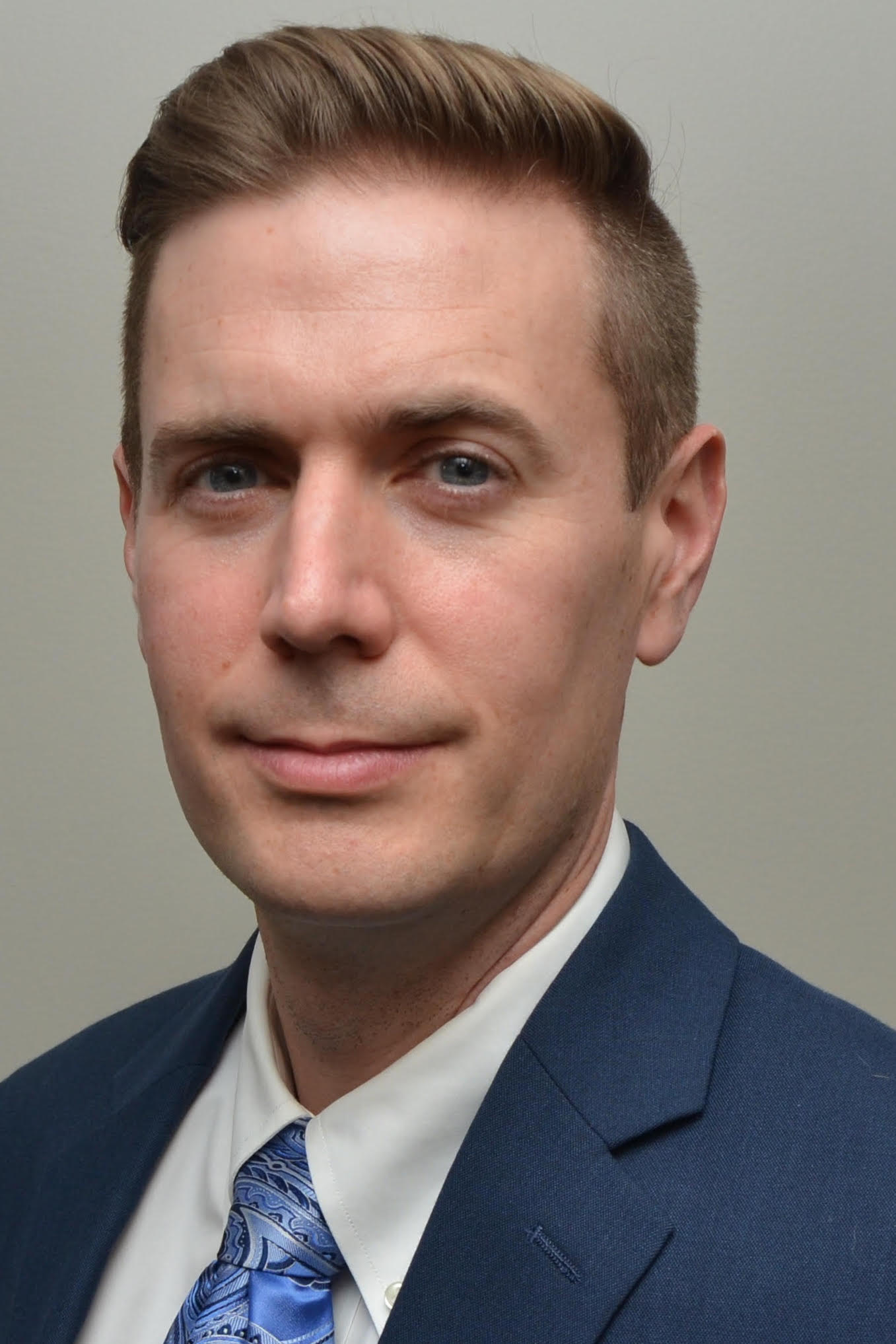}}]{David J. Love}
 (S’98 - M’05 - SM’09 - F’15) received the B.S. (with highest honors), M.S.E., and Ph.D. degrees in electrical engineering from the University of Texas at Austin in 2000, 2002, and 2004, respectively. Since 2004, he has been with the Elmore Family School of Electrical and Computer Engineering at Purdue University, where he is now the Nick Trbovich Professor of Electrical and Computer Engineering. He served as a Senior Editor for IEEE Signal Processing Magazine, Editor for the IEEE Transactions on Communications, Associate Editor for the IEEE Transactions on Signal Processing, and guest editor for special issues of the IEEE Journal on Selected Areas in Communications and the EURASIP Journal on Wireless Communications and Networking. He was a member of the Executive Committee for the National Spectrum Consortium. He holds 32 issued U.S. patents. His research interests are in the design and analysis of broadband wireless communication systems, beyond-5G wireless systems, multiple-input multiple-output (MIMO) communications, millimeter wave wireless, software defined radios and wireless networks, coding theory, and MIMO array processing.

Dr. Love was named a Thomson Reuters Highly Cited Researcher (2014 and 2015), is a Fellow of the Royal Statistical Society, and has been inducted into Tau Beta Pi and Eta Kappa Nu. Along with his co-authors, he won best paper awards from the IEEE Communications Society (2016 Stephen O. Rice Prize and 2020 Fred W. Ellersick Prize), the IEEE Signal Processing Society (2015 IEEE Signal Processing Society Best Paper Award), and the IEEE Vehicular Technology Society (2010 Jack Neubauer Memorial Award).
\end{IEEEbiography}
\begin{IEEEbiography}[{\includegraphics[width=1in,height=1.25in,clip,keepaspectratio]{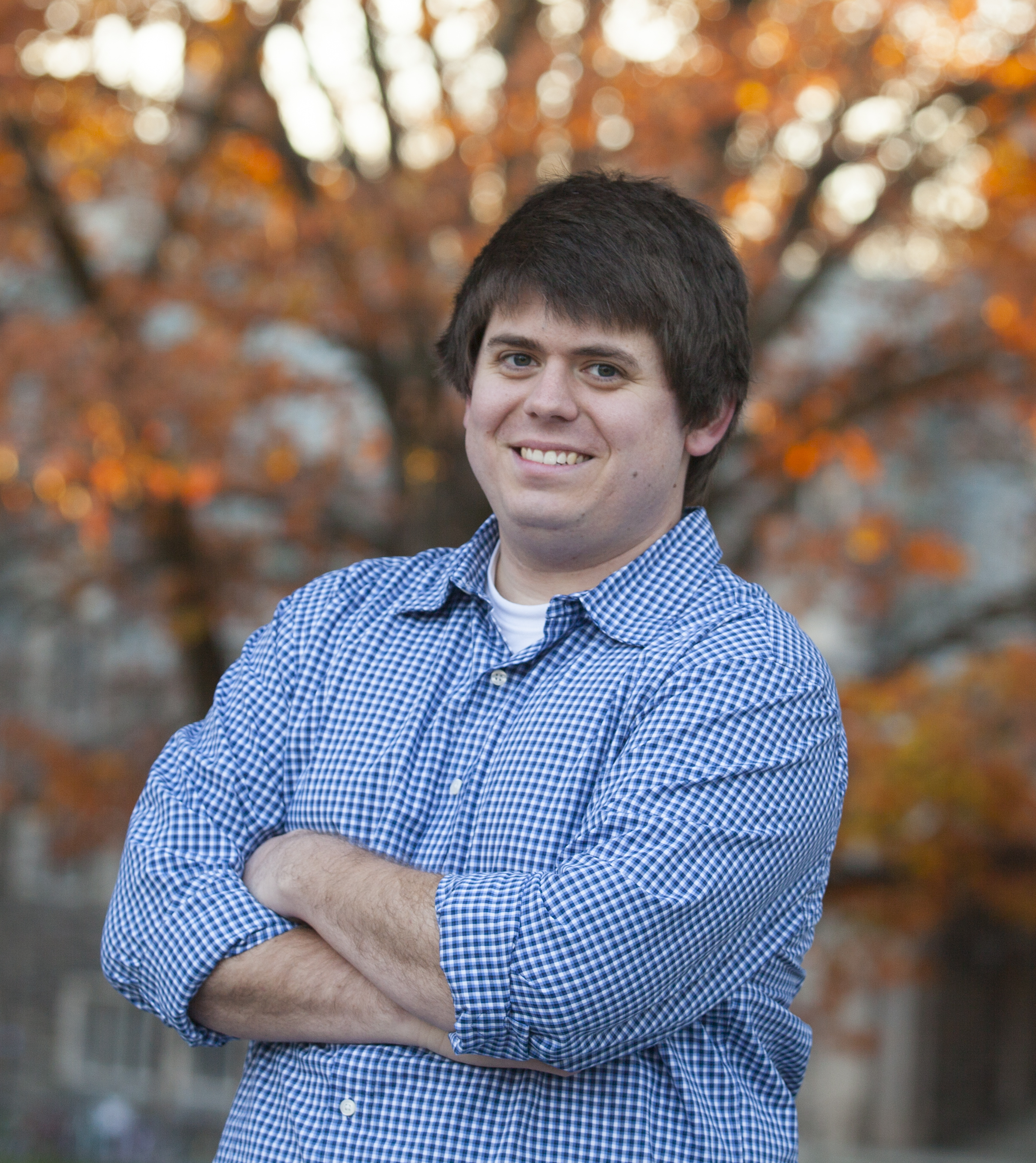}}]{Christopher G. Brinton}
 (S’08, M’16, SM’20) is an Assistant Professor in the School of Electrical and Computer Engineering at Purdue University. His research interest is at the intersection of networked systems and machine learning, specifically in distributed machine learning, fog/edge network intelligence, and data-driven network optimization. Dr. Brinton was the recipient of the 2022 NSF CAREER Award, 2022 ONR Young Investigator Program (YIP) Award, and 2022 DARPA Young Faculty Award (YFA). He currently serves as an Associate Editor for IEEE Transactions on Wireless Communications, in the ML and AI for wireless area. Prior to joining Purdue, Dr. Brinton was the Associate Director of the EDGE Lab and a Lecturer of Electrical Engineering at Princeton University. Dr. Brinton received the PhD (with honors) and MS Degrees from Princeton in 2016 and 2013, respectively, both in Electrical Engineering.
\end{IEEEbiography}

\bibliographystyle{IEEEtran}
\bibliography{refs}

\end{document}